\documentclass[10pt,journal,compsoc]{IEEEtran}
\usepackage{caption}
\DeclareCaptionType{copyrightbox}

\usepackage{array}
\usepackage{multirow, flushend}
\usepackage{subfig}
\usepackage[final]{graphicx}
\usepackage{amsmath}
\usepackage{pbox}
\usepackage{xcolor}
\usepackage{footmisc}
\usepackage{amssymb}

\ifCLASSOPTIONcompsoc
  \usepackage[nocompress]{cite}
\else
  \usepackage{cite}
\fi
\ifCLASSINFOpdf
\else
\fi
\usepackage{placeins}

\usepackage{array}
\usepackage{paralist}
\newcommand{\subparagraph}{}
\usepackage[compact]{titlesec}
\titlespacing{\section}{0pt}{1ex}{0ex}
\titlespacing{\subsection}{0pt}{1ex}{0ex}
\titlespacing{\subsubsection}{0pt}{0.5ex}{0ex}
\def\@IEEEauthorblockAstyle{\smallfont\normalsize}

\begin{document}
%
\title{Exploration of Performance and Energy Trade-offs for Heterogeneous Multicore Architectures}
%
%
%

\author{
\IEEEauthorblockN{
Anastasiia~Butko,
		Florent~Bruguier,
        David Novo,
        Abdoulaye~Gamati\'{e},
        Gilles~Sassatelli } \\
LIRMM (CNRS and University of Montpellier), Montpellier, France\\
\{firstname.lastname\}@lirmm.fr
}

\IEEEtitleabstractindextext{%
\begin{abstract}
Energy-efficiency has become a major challenge in modern computer systems. To address this challenge, candidate systems increasingly integrate heterogeneous cores in order to satisfy diverse computation requirements by selecting cores with suitable features. In particular, single-ISA heterogeneous multicore processors such as ARM big.LITTLE have become very attractive since they offer good opportunities in terms of performance and power consumption trade-off. The key design principle of these processors relies on the combination of low-power in-order cores with high-performance out-of-order cores. While existing works already showed that this feature can improve system energy-efficiency, further gains are possible by generalizing the principle to higher levels of heterogeneity. 
The present paper aims to explore these gains by considering single-ISA heterogeneous multicore architectures including three different types of cores. For this purpose, we use the Samsung Exynos Octa 5422 chip as baseline architecture. Then, we model and evaluate Cortex A7, A9, and A15 cores using the gem5 cycle-approximate simulation framework coupled to McPAT for power-consumption estimation. A thorough design space analysis is carried out considering the Rodinia benchmark suite. We demonstrate that varying the level of heterogeneity as well as the different core ratio can lead to up to 2.3x gains in energy efficiency and up to 1.5x in performance.
This study further provides insights on the impact of workload nature (e.g., level of parallelism, computation complexity, communication pattern) on performance/energy trade-off and draws recommendations concerning suitable architecture configurations.
This contributes \textit{in fine} to guide future research towards dynamically reconfigurable HSAs i.e. architectures in which some cores / clusters can be disabled momentarily so as to optimize certain metrics such as energy efficiency. This is of particular interest when dealing with quality-tunable algorithms in which accuracy can be then traded for compute effort, thereby enabling to use only those cores that provide the best energy-efficiency for the chosen algorithm.

\end{abstract}

}

\maketitle

\IEEEdisplaynontitleabstractindextext

%
\IEEEpeerreviewmaketitle

%
%
%
%

\IEEEraisesectionheading{\section{Introduction}
\label{sec:introduction}}

In many computation contexts, finding a good compromise between the quality of calculated results and computation complexity is important. Complexity often drives performance requirements. For example, in video processing various motion estimation algorithms are applicable with tunable rendering quality, within reasonable execution time. In fact, optimal motion estimation algorithms, e.g., Full Search algorithms \cite{Monteiro:2014:PFS:2589752.2589763}, usually involve intensive computations that are time consuming. On the contrary, sub-optimal algorithms, e.g., SLIMPEG \cite{Alfonso2002UltraLM}, can be executed in a very reasonable time while the quality of renderings remains acceptable enough. Another example concerns heuristic algorithms, which generally address complex problems in shorter times compared to optimal algorithms, at the cost of low quality results, yet relevant enough in the scope of the considered application context. Relaxing the need for fully precise results therefore helps reducing the computation load, hence enables a faster algorithm execution.

Beyond  execution time, another motivation behind quality degradation via computationally reasonable algorithms is energy-efficiency improvement. In order to accommodate the computation requirements of such algorithms w.r.t. power consumption of the underlying compute systems, heterogeneous architectures offer an interesting opportunity. 
Indeed, with the ever-increasing 
demand in system energy-efficiency, 
a strong focus has been put 
on new system designs that can potentially extend digital performance scaling and improve power consumption \cite{7368023}. In contrast to general-purpose computers leading the computing field for many years, 
future embedded systems such as advanced driver-assistance systems, tablets or smartphones, which integrate a wide range of functionalities/applications, 
will have an extreme level of heterogeneity, combining a variety of specialized compute accelerators with general-purpose processors in order to satisfy diverse performance requirements.
This approach, usually referred to as Heterogeneous System Architecture (HSA), has already been adopted by industry resulting in a well-known graphic acceleration through the GPU technology and more recent architectures dedicated to AI / Machine Learning, such as Nervana's AI platform~\cite{DBLP:journals/corr/Lavin15b} and Google TPU~\cite{Jouppi:2017:IPA:3079856.3080246}.  
More generally, given some target performance levels, suitable cores can be selected within a HSA system for computation, so as to achieve the best compromises in terms of result quality, performance and energy\footnote{Note that this goal is also shared by the recent approximate computing paradigm \cite{DBLP:conf/ets/HanO13}.}. 

However, potential HSA benefits come at the cost of severe programming complexity. In case of GPU programming, the complexity basically comes from the memory hierarchy. Space allocation and data movement tasks are not automated, and some efforts \cite{Ueng2008} are yet entrusted to the programmer. 

Single-ISA heterogeneous multicore systems~\cite{Kumar} offer an attractive alternative. While providing a certain level of heterogeneity through core microarchitecture difference, cores share the same ISA and feature binary compatibility. Thus, the programming environment does not differ from the conventional methods, and most of the complexity lies in efficient task scheduling. 
In particular, this technology gains popularity in the mobile market~\cite{Samsung:Exynos}\cite{qualcomm}\cite{Nvidia:Tegra}, where energy-efficiency usually prevails over high performance. The task scheduler runs popular applications such as web browsing or music playback on fast cores while simultaneously maintaining background tasks on slower low-power cores. Thus, a certain performance/power balance can be achieved that helps extending battery life.  Within a single parallel workload, the task scheduling problem lies in task-communication graph analysis and partitioning to decide which tasks have to be scheduled on the high-performance cores, which ones on the low-power cores and at which level of granularity \cite{ChronakiRBALV15}\cite{openmp}.

Existing work \cite{VanCraeynest:2013:UFD:2400682.2400691} already tried to answer basic design questions such as how many different core types should be integrated into processor, how they should communicate among each other, etc. On the other hand, each workload may require a different architecture configuration. This motivates further research towards dynamically reconfigurable HSAs. Besides the extensive software support, such HSAs require a deep understanding of what are the application characteristics that require heterogeneous architecture to be dynamically reconfigured and how to design the corresponding architecture. Finally, it is important to have an assessment of the benefits provided by such a reconfiguration in terms of performance and energy, as well as the related reconfiguration overheads. Answering the above questions strongly requires a design space exploration. 
As the complexity of HSAs introduces architecture  modeling challenges, existing attempts to address them use dynamic voltage/frequency scaling (DVFS) \cite{Takouna_efficientvirtual} to emulate different cores or rely on high-level analytical approaches \cite{VanCraeynest:2013:UFD:2400682.2400691}. In both cases, realistic architecture modeling 
cannot be guaranteed \cite{Koufaty:2010:BSH:1755913.1755928}. Since application nature has a strong impact on the architecture and vice versa, the exploration methodology requires a software-hardware co-simulation approach.

\textbf{Objective and contribution of this paper.}
The contributions of this paper can be summarized as follows:

\begin{itemize}
\item Study of different HSA configurations for better energy-performance trade-offs. In particular, we analyze the gains resulting from increasing the level of heterogeneity to three and allowing for \textit{asymmetric heterogeneous } configurations. We then identify on the chosen benchmark suite the best performing configurations according to several performance metrics.

\item Correlation between obtained results and benchmark nature according to various characteristics such as memory pressure and arithmetic complexity, so as to draw generic guidelines for HSA definition.

\end{itemize}

The insights gained from this study can be leveraged for the energy-efficient execution of quality-tunable algorithms, by carefully selecting the most suitable cores within a multicore heterogeneous system. This can be further extended to modifying at run-time the logical HSA system by means of disabling cores or even clusters for the entire duration of the said quality-driven algorithm configuration. This makes for further gains in energy efficiency as part of the system does not participate in the power consumption, in a \textit{dark silicon} fashion.

\begin{table*}[t]
\centering
\footnotesize
\caption{ARM Cortex-A processor core microarchitectures}
\begin{tabular}{l|c|c|c|c} 
\textbf{Feature} & \textbf{Cortex-A7} & \textbf{Cortex-A9} & \textbf{Cortex-A15} &\textbf{Comment} \\  \hline \hline
\textit{Out-of-Order support} & \centering No & Yes & Yes & -  \\ [2pt]
\textit{Decoders}  &2 & \pbox{4cm}{\centering 2} & 3 & \pbox{6cm}{\centering Extensive use of shared memory}\\ [2pt]
\textit{Pipeline stages} & 8 & 9 & 15 & -\\ [2pt]
\textit{Execution units} & 5 & 5 & 8 & \pbox{4cm}{\centering -}\\ [2pt]
\textit{FPU}  & VFPv4 & VFPv3  & VFPv4 & \pbox{6cm}{\centering VFPv4 supports Fused Multiply Add (FMA)} \\ [2pt]
\textit{L1I / L1D private caches}  &32kB & 32kB & 32kB & \pbox{6cm}{ \centering Design-time configurable size} \\ [2pt]
\textit{L2 shared cache}  & 512kB & - & 2MB & \pbox{9cm}{\centering Configurable sizes (sizes for Exynos 5422) and shared at cluster-level } \\
\textit{AXI bus interfaces}  & 1 x 128-bit & 2 x 64-bit & 1 x 128-bit & \pbox{6cm}{\centering - } \\

\hline \hline
\end{tabular}
\label{tab:cortex_specifics}
\end{table*}

In order to address HSA design issue w.r.t. aforementioned questions, we consider the gem5 cycle-approximate simulation framework \cite{Binkert:2011:GS:2024716.2024718} coupled with McPAT tool \cite{McPAT} for respectively architecture modeling / full-system workload execution and energy efficiency assessment. The baseline architecture model, i.e., ARM big.LITTLE processor, has been previously validated against the real hardware \cite{mcsoc} to enable realistic performance and energy estimation. We profile a target set of scientific workloads from the \texttt{Rodinia} benchmark suite \cite{5306797} on the Exynos 5 Octa chip to assess the impact of the application nature on HSA system execution. Using our modeling environment, we evaluate application performance and energy on different heterogeneous architectures. 

\textbf{Outline.} In the rest of the paper, Section \ref{sec:background} provides background on existing single-ISA heterogeneous System-on-Chips (SoCs), their designs, commercial trends and programmability challenges, as well as related work. Section \ref{sec:methodology} describes our methodology for architecture evaluation and benchmarking. In Section \ref{sec:analysis}, we provide a detailed analysis of chosen scientific workloads. Section \ref{sec:results} shows the results of the architecture exploration and provides insights. Finally, Section \ref{sec:conclusions} points out conclusions and future work.

\section{Background}
\label{sec:background}
 
Here, we provide a short overview on the existing single-ISA heterogeneous SoCs, their designs, commercial trends and programmability challenges.

\subsection{Design choice}
\label{subsec:21}
Over the past five years, single-ISA heterogeneous SoCs have significantly gained in popularity due to the increased attention from industrial leaders and their continuing contribution to the field. 
The early NVIDIA's SoCs, i.e. Tegra 3 and 4, represent Variable Symmetric Multiprocessing (vSMP) technology that combines four faster power-hungry cores together with one `companion' core dedicated to background tasks. All five cores have similar architecture, but the main cores are built in a standard silicon process to reach higher frequencies and the `companion' core is built using a special low power silicon process that executes tasks at a lower frequency \cite{nvidiawp}. 

Most existing processors implement the ARM big.LITTLE architecture that differs from the vSMP in the concept of `companion' core providing instead a combination of `big'  and 'LITTLE' cores. These cores are built using the same process, but because of the difference in their micro-architecture, they have diverse performance/power characteristics. Also, `big'  and 'LITTLE' cores are combined in such a way as to form \textit{symmetric} configurations, e.g. HiSilicon K3V3, Samsung Exynos 5/7 Octa, Nvidia Tegra X1, or \textit{asymmetric},  e.g. HiSilicon Kirin 920, Samsung Exynos 8 Octa. In contrast to the vSMP approach, current asymmetric configurations rather contain a smaller number of big cores and a higher number of LITTLE cores. Such configurations are often suggested in the literature as in \cite{dal}, in which authors promote a sequential accelerator associated with several simpler cores (for parallel code regions) as a very attractive design solution. In the rest of the paper, \textit{symmetric} and \textit{asymmetric} configurations always denote system configurations such that the number of cores in each cluster is \textit{identical} and \textit{different} respectively.

%
%
%
%
%
%

Since Cortex-A15 has been released in 2012, it became the main workhorse in most listed SoCs due to its high performance and reasonable power consumption. For low-power tasks, Cortex-A15 processor cores are usually paired with Cortex-A7 cores. ARM reports over 50\%  energy savings for popular activities, such as web browsing and music playback, with such Cortex-A7/Cortex-A15 configuration \cite{armwp}. With the advent of the new 64-bit ARMv8 ISA, Cortex-A15 and Cortex-A7 are substituted by Cortex-A57 and Cortex-A53 as big and LITTLE cores respectively. Moreover, recently released MediaTek Helio X20/25 SoCs contain three clusters instead of two pushing the level of heterogeneity. In terms of microarchitecture, Helio X20/25 combines two types of core, Cortex-A57 and Cortex-A53. Though sharing the same Cortex-A53 microarchitecture, Medium and Little clusters use different technology library implementations with different power/performance profiles, resulting in different maximum frequencies.



Besides the performance and power characteristics, the area constraint also plays a significant role in the design choice. Indeed, typical 28~nm silicon footprint for the Cortex-A7 core is 0.45 $mm^2$ that is almost seven times smaller than the 3.1 $mm^2$ footprint of the Cortex-A15 core. This usually advocates for design decisions towards limited big core count, paired with smaller cores as the latter come at much lower cost area-wise.
Moreover, silicon process and performance scaling issues motivate research to revisit basic architectural concepts for more efficient use of available silicon area within a given power envelope. Thus, \textit{Single-ISA heterogeneous architectures} capable of supporting run-time reconfigurations based on specific application requirements constitute a promising line of research. 

Several approaches address such dynamic heterogeneity.
The first group of works, e.g. Core Fusion \cite{Ipek:2007:CFA:1273440.1250686}, TFlex \cite{Kim:2007:CLP:1331699.1331733}, WiDGET \cite{Watanabe:2010:WWD:1816038.1815965}, proposes to build the architecture with multiple small cores for parallel workloads with high thread-level parallelism (TLP). These cores can be dynamically `fused' into a large high-performance core for sequential code sections with the instruction-level parallelism (ILP). An opposite approach implies core micro-architecture adaptation, i.e. an adaptive out-of-order core is used as a basic unit. It can be dynamically transformed into a highly-threaded in-order core. There are several implementations of this approach, such as MorphCore \cite{6493629}, FLicker \cite{Petrica:2013:FDA:2508148.2485924},  ElasticCore \cite{Tavana:2015:EED:2744769.2744833}. Yet, none of the proposed approaches has been adopted by industry to become commercially available. The reason lies in their programmability that in addition to the problem of mapping computational tasks on appropriate processing units, has to deal with reconfiguration overhead.

\subsection{Programmability and task scheduling}

Existing programming approaches for heterogeneous big.LITTLE-like architectures are usually classified into three execution modes \cite{arm_software}. 
%
%
%
%

The first and simplest mode is \textit{cluster migration}. A single cluster is active at a time, and migration is triggered on a given workload threshold. This mode provides low flexibility in switching granularity and in power saving opportunities. Maximum available performance corresponds to fully used `big' cluster, i.e. N `big' cores.  The second mode named \textit{core migration} relies on pairing every `big' core with a `LITTLE' core. Each pair of cores acts as a virtual core in which only one actual core among the combined two is powered up and running at a time. Only four physical cores at most are active. The main difference between clustered migration and core migration modes is that the four actual cores running at a time are identical in the former while they can be different in the latter. This mode provides average flexibility in switching and power saving. Maximum available performance is similar to the cluster migration mode and corresponds to fully used `big' cluster.  The \textit{heterogeneous multi-processing mode} (HMP) implies using all the cores together. A strong argument in favor of HMP is that it provides a fine-grained control of workloads and consequently opens a possibility for high level of power saving. Maximum available performance is higher compared to the cluster migration and core migration modes and corresponds to N `big' cores and M `LITTLE' cores running simultaneously. Thus, the most promising and yet little studied HMP execution mode is our target research in this work.

Exploiting the potential of single-ISA heterogeneous multicore architectures in HMP mode requires appropriate strategies to manage the distribution of computational tasks among different cores, and is tightly coupled to the used programming model.

OpenMP \cite{openmp} is a popular shared memory parallel programming interface. OpenMP features thread-based fork-join task allocation model. It consists of a set of compiler directives, library routines and environment variables for developing parallel applications. The OpenMP loop scheduling allows determining the way in which iterations of a parallel loop are assigned to threads. Iterations can be assigned in \textit{chunks}, e.g. the number of contiguous iterations.



There are several alternative programming models and scheduling policies dedicated to single-ISA HSAs. Yu et al. in \cite{6864009} evaluate ARM big.LITTLE power-aware task scheduling, via power saving techniques such as dynamic voltage and frequency scaling (DVFS) and dynamic hot plug. Tan et al. in \cite{7059077} implement a computation approximation-aware scheduling framework in order to minimize energy consumption and maximize quality of service, while preserving performance and thermal design power constraints. 
Our approach follows a different route and relies on a standard OpenMP implementation; focus is put on relating application profiles to architecture characteristics from performance and energy-efficiency perspectives. We further decide to use dynamic scheduling policy, which is motivated by a previous study that has shown only this scheduler is able to exploit architecture heterogeneity\cite{butko2017efficient}. 




\subsection{Heterogeneous system architecture modeling}
\label{subsec:23}
The design choice for both statically and dynamically heterogeneous architectures conceals many challenges. To achieve the optimal trade-off between system performance and energy, an extensive design space exploration is required. The per-program performance/energy profile is rather difficult to predict, especially given the broad range of workload characteristics, such as computation complexity, data movement intensity, arithmetic complexity etc. A large part of studies on single-ISA HSAs focuses on such evaluations. In \cite{VanCraeynest:2013:UFD:2400682.2400691} authors aim at providing some fundamental design insights based on a high-level analytical model analysis. Particularly, they claim two core types being the most beneficial configuration and a higher level of heterogeneity does not contribute much. Endo et al. \cite{Endo:1,Endo:2} show the micro-architectural simulation of ARM Cortex-A cores of the big.LITTLE processor by using the gem5/McPAT frameworks and validate area and energy/performance trade-offs against the published datasheet information. Authors target single-core evaluation only, neither heterogeneous multicore architectures, nor parallel application execution are covered by their work. 

In addition to existing modeling works, to answer the design space exploration questions described in Section \ref{sec:introduction}, an advanced comparison and calibration against a real platform is strongly required for ensuring the consistency of reported simulation results. 
Our work advances state-of-the-art in heterogeneous system modeling by providing joint application-architecture exploration. We use previously validated models \cite{mcsoc} of the big.LITTLE multicore system to evaluate the trade-offs in terms of performance delay and energy, and provide a detailed analysis of the chosen workloads.

\begin{figure}[t]
\centering
\includegraphics[scale=0.5]{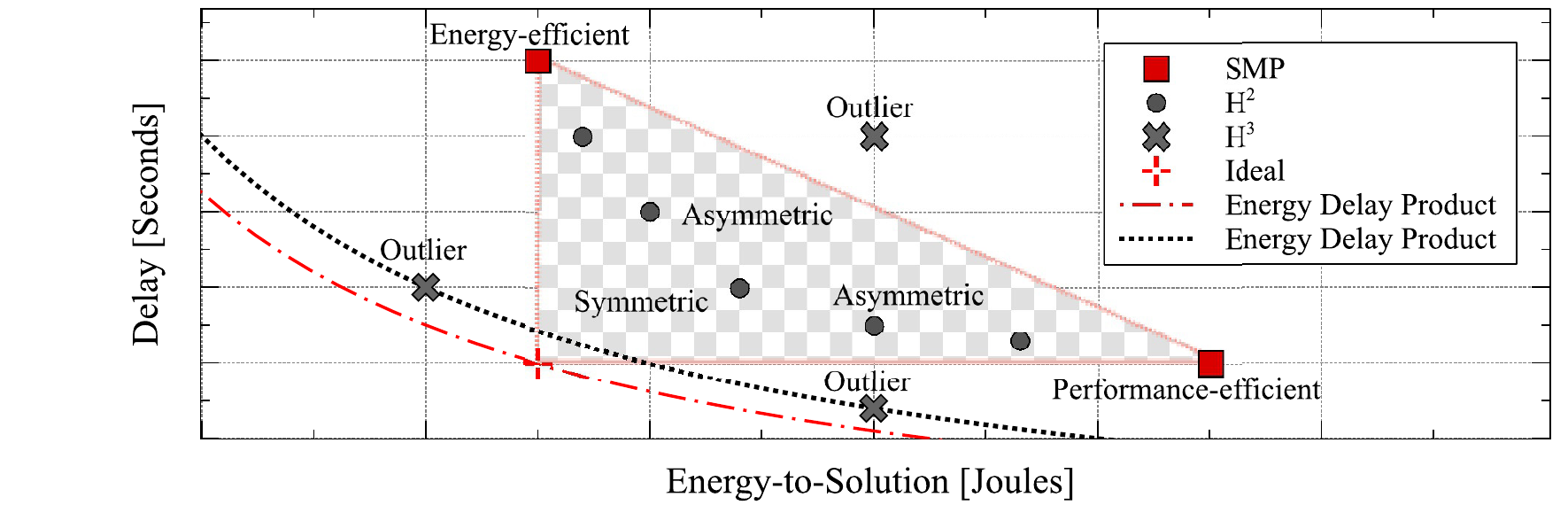}
\caption{Delay, EtoS and EDP projection.}
\label{fig:projection}
\end{figure}

\section{Methodology}

\label{sec:methodology}

\subsection{Baseline architecture}
Our baseline architecture is the Samsung Exynos 5 Octa (5422) multicore chip \cite{Samsung:Exynos}. The processor features two heterogeneous clusters, ``big'' and ``LITTLE'', each of which consists of four Cortex-A15 and four Cortex-A7 cores respectively. Clusters operate at independent frequencies, from 200MHz up to 1.4GHz for the LITTLE and up to 2GHz for the big. Table \ref{tab:cortex_specifics} gives the specifics of the Cortex-A7 and Cortex-A15 cores as of implemented in this SoC, alongside Cortex-A9 which is used later on in the exploration section.




\begin{figure}[t]
  \centering
    \includegraphics[width=\linewidth]{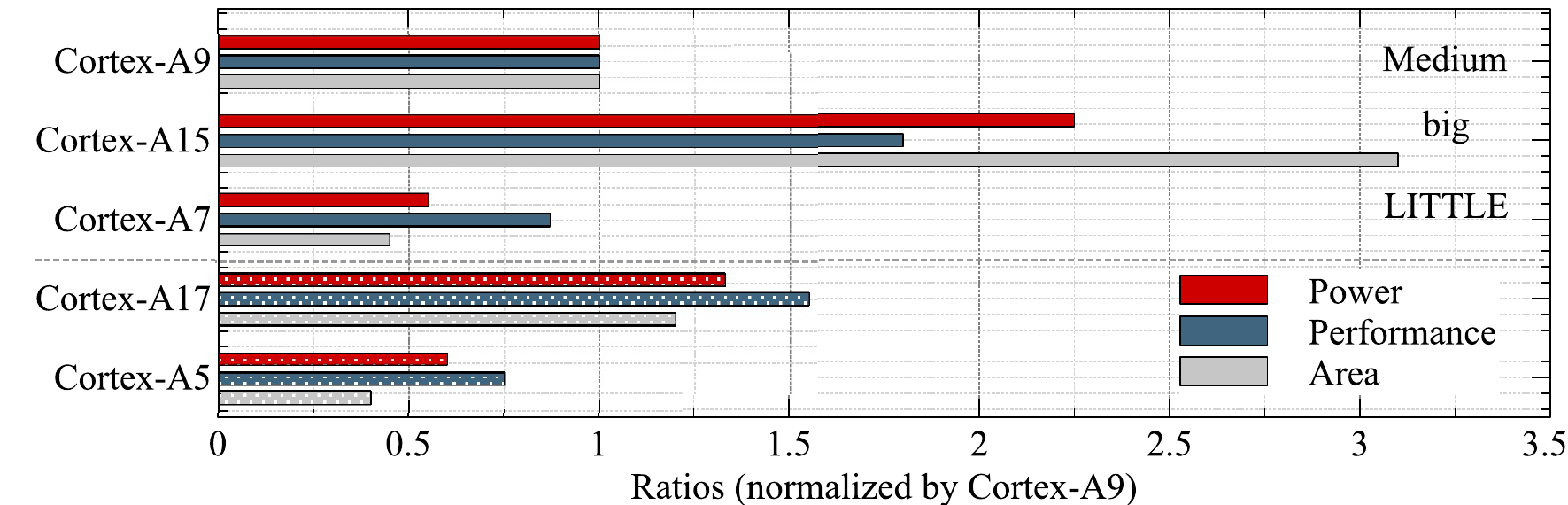}
  \caption[ARM Cortex-A series performance/power ratios]{ARM Cortex-A series performance/power ratios.}
  \label{fig:ratio}
%
%
\end{figure}

\begin{table*}[!ht]
\centering
\footnotesize
\caption{Detailed \texttt{Rodinia} Benchmark analysis.}
\begin{tabular}{l|c|c|c|c} 
\textbf{Workload} & \textbf{Problem size} & \textbf{Dwarve} &\textbf{Bottlenecks} & \textbf{Features}\\  \hline \hline
\textit{heartwall} & \centering test.avi & Structured & Memory Bandwidth & Braided parallelism (task and data). \\ [5pt]
\textit{lud}  &2k & Dense Linear & Computation & \pbox{7cm}{ \centering Inter-thread row and column dependencies. Extensive use of shared memory.} \\ [10pt]
\textit{nw}  &1k & \pbox{4cm}{\centering Dynamic Programming} & Memory Latency & \pbox{7cm}{\centering Extensive use of shared memory.}\\ [5pt]
\textit{kmeans} & 200k & Dense Linear & Computation & Massive data parallelism.\\ [5pt]
\textit{nn} & 42k & Dense Linear & Computation & \pbox{4cm}{\centering Control flow divergence.}\\ [5pt]
\textit{backprop}  & 64k & Unstructured  & Memory Latency & \pbox{5cm}{\centering Extensive use of shared memory.} \\ [5pt]
\textit{srad v1}  &1x502x458 & Structured & Memory Bandwidth & \pbox{6cm}{\centering Synchronization dependencies between stages. High intra-thread data locality.}\\
\hline \hline
\end{tabular}
\label{tab:problem_size}
\end{table*}

\subsection{Design space exploration}
\label{subsec:32}
We explore the design space in terms of performance and energy trade-offs. A set of evaluation metrics for this study contains the \textit{Execution Time} or \textit{Delay}, \textit{Energy-to-Solution (EtoS)} and \textit{Energy Delay Product (EDP)}.

Figure \ref{fig:projection} shows the projection of Delay, EtoS and EDP that we aim to study for different HSA configurations. The first group of points depicted as red squares presents homogeneous or symmetric multicore processors (SMP), i.e. the most energy-efficient and the most performance-efficient. In the Ideal scenario -- red cross, both benefits are achieved. Dash-dotted red line shows the points that provide the EDP equal to the ideal scenario while giving different performance/energy values. 

%
%

The strategy of combining such two opposite types of core in a single HSA processor ($H^2$) promises to get closer to the Ideal scenario. Varying the ratio of core types, i.e. symmetric or asymmetric configurations, a priority to one target metric may be given. The checkerboard triangle outlines the potential scope for such heterogeneous configurations. 

To go beyond, we consider three-level heterogeneous processors ($H^3$) with three types of cores. We expect that these configurations provide outliers with even better Delay/EtoS trade-off and EDP.  Dotted black line shows the EDP curve provided by the outliers.

Here, we define three core types that are chosen to configure the evaluated processor architectures. big and LITTLE cores are the same as in the baseline architecture, i.e. ARM Cortex-A15 and ARM Cortex-A7 respectively. We also include the third core type called Medium that represents ARM Cortex-A9 core. Our choice is based on the performance, power consumption and area ratios shown in Figure \ref{fig:ratio}. These ratios are normalized against Cortex-A9 and represent averaged values among multiple calculation types (e.g. floating-point, integer)\footnote{Shown ratios are based on the publicly available information   https://developer.arm.com/products/processors/cortex-a/\label{fnlabel2}}. Among the existing ARM Cortex-A series such as Cortex-A5 and Cortex-A17, the combination of Cortex-A15, -A9 and -A7 provides suitable compromise to configure potentially balanced HSA.    

The design space contains three architecture groups. The first group represents SMP processors with only big, Medium or LITTLE cores. The second group contains heterogeneous architectures with two-level granularity, i.e. $H^2$: base architecture with symmetric core types configuration, i.e. 4A7/4A15, and asymmetric configurations, such as 7A7/1A15 and 7A9/1A15. The third group represents heterogeneous architectures with three-level granularity, i.e. $H^3$, that we name big.Medium.LITTLE architecture. We vary the number of cores per cluster going from 1A7/1A9/6A15 to 1A7/6A9/1A15 and then to 6A7/1A9/1A15.

Listed architecture configurations have a common structure: each cluster shares one L2 cache of 1MB, core frequency is 1GHz and the total number of cores per system is 8. In this way, we negate the impact of these parameters on the results and evaluate only the effect of different core types.  

\begin{figure*}[t]
\centering
\subfloat[Functional Summary]{\includegraphics[scale=0.49]{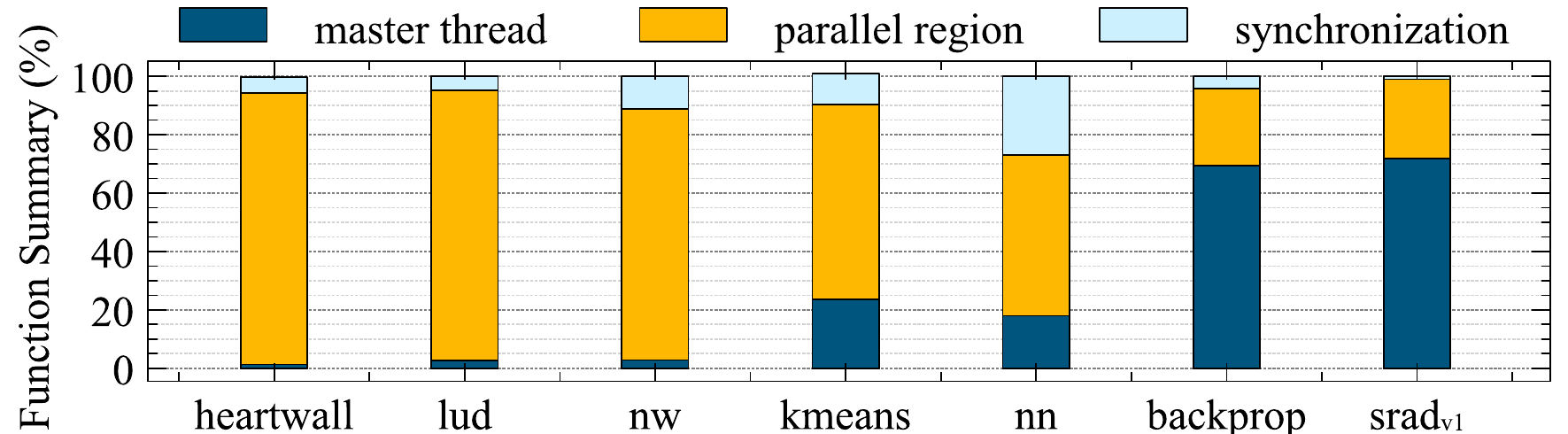}}
\hspace{5pt}
\subfloat[Instruction Summary]{\includegraphics[scale=0.49]{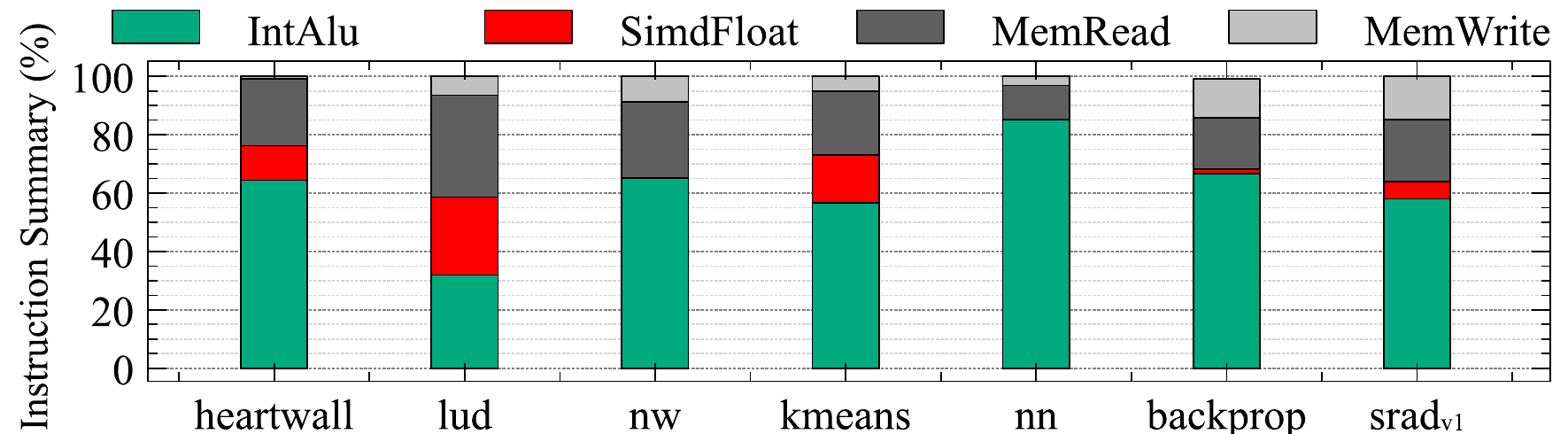}}
\caption{Rodinia Workloads Analysis.}
\label{fig:summary}
\end{figure*}


\subsection{Full-system Simulation}
We use the gem5 simulator which is an event-driven cycle-approximate simulator \cite{Binkert:2011:GS:2024716.2024718}. It
has a modular structure that enables flexible configuration of various
multicore architecture components. gem5 supports several simulation modes,
which differ in speed and accuracy. Namely, the \textit{Syscall Emulation
(SE)} mode emulates most operating system services and devices using stubs on
the host computer. It achieves high simulation speed at the cost of limited
accuracy. On the other hand, \textit{Full-System (FS)} mode is able to run an
unmodified operating system, as if this was running on the real hardware.
Although highly accurate, the FS mode incurs in significantly longer simulation
times. gem5 provides a large set of architecture design components including multiple
ISAs, CPU models and memory systems. It further produces statistical information enabling to estimate power consumption and footprint area with the Multicore Power, Area, and Timing (McPAT) modeling framework \cite{McPAT}.

We implemented performance and power simulation models of ARM big.LITTLE multicore processor. Our recent study evaluates the accuracy of the proposed models against the real Samsung Exynos Octa (5422) SoC \cite{mcsoc}. On an average, the gem5 model predicts performance with less than 20\% error. The average error percentage of total power is around 12\%.

\section{Workloads Profiling}
\label{sec:analysis}
The study is conducted using the \texttt{Rodinia} benchmark suite \cite{5306797}. It is composed of applications and kernels from different domains such as bioinformatics, image processing, data mining, medical imaging and physics simulation. \texttt{Rodinia} targets performance benchmarking of heterogeneous systems, and the benchmark exists in CUDA, OpenMP and OpenCL implementations. Out of the \texttt{Rodinia} benchmark suite, we select the following subset of benchmarks: \textit{Back Propagation}, \textit{Heart Wall}, \textit{kmeans openmp}, \textit{lud}, \textit{nn}, \textit{nw} and \textit{srad v1}. Based on dwarfs classification proposed in \cite{Asanovic06thelandscape}, the chosen subset contains Structured Grid (\textit{heartwall}, \textit{srad}), Unstructured Grid (\textit{backprop}), Dynamic Programming (\textit{nw}), Dense Linear Algebra (\textit{kmeans}, \textit{lud}, \textit{nn}) dwarfs.


To profile chosen workloads, we use the Odroid XU3 board \cite{Hardkernel} built around the Exynos 5 Octa (5422) SoC. The board runs Ubuntu 14.04 OS on Linux kernel LTS 3.10. Here, the OpenMP implementation of \texttt{Rodinia} benchmark is chosen, with four threads per cluster, i.e., one thread per core\footnote{None of the evaluated ARM cores support simultaneous multi-threading.\label{fnlabel1}}. We use \textit{dynamic} loop scheduling policy with chunk size equal to `1' such that maximum runtime freedom is given. Threads are bound to specific CPU cores by using \texttt{GOMP\_CPU\_AFFINITY} environment variable in a such way to ensure that the main thread is executed on the big core. 

The analysis is conducted using the Scalasca/Score-P instrumentation \cite{Scalasca} and Vampir event trace data visualization tool \cite{Vampir}. Scalasca is an open-source toolset that allows analyzing parallel application execution behavior. Through the source code instrumentation, application is elaborated with specific directives that notify the measurement library of performance-relevant runtime events whenever they occur. The runtime overhead varies between 1.3\% and 30\% depending on application and architecture configurations \cite{scalrep}. Scalasca is designed to identify potential performance bottlenecks in particular those concerning communication and synchronization. Additional information concerning application computation patterns is obtained by full-system simulation and subsequent per-core statistic analysis.

\begin{figure}[ht]
  \centering
    \includegraphics[scale=0.5]{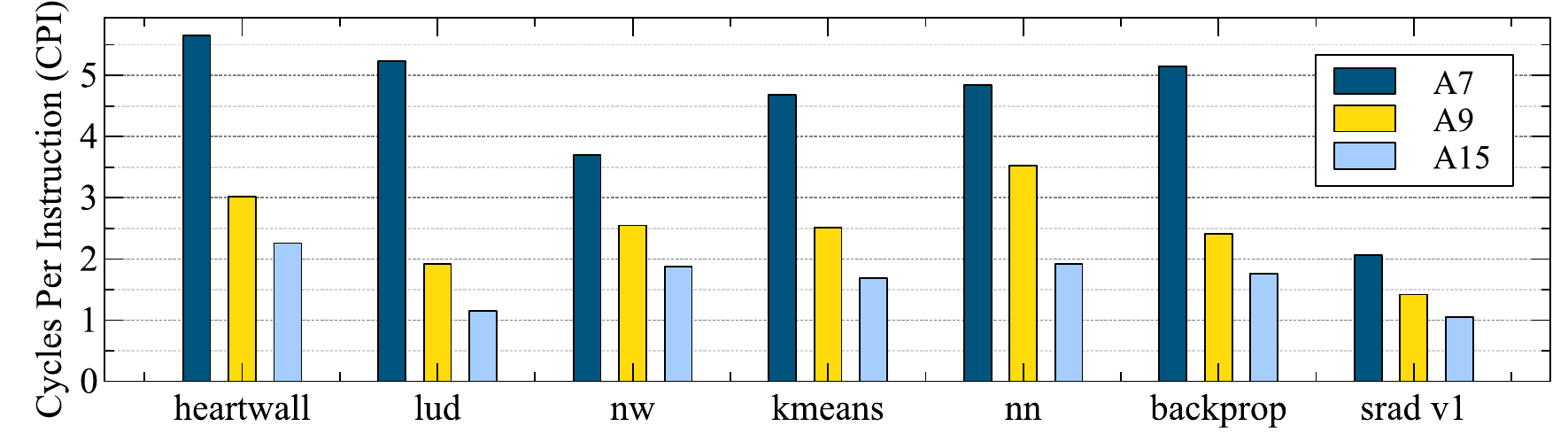}
  \caption[ARM Cortex-A series performance/power ratios]{\textit{Cortex-A series CPI comparison.}}
  \label{fig:cpi}
\end{figure}

Table \ref{tab:problem_size} summarizes the main characteristics of the chosen workloads.
Besides general characteristics that are well-defined according to the dwarf classification, our analysis includes detailed functional and instruction summary. Functional summary shown in Figure \ref{fig:summary} (a) represents the percentage of time spent in different stages, such as master thread execution, parallel region and synchronization. 
It has been obtained by applying the Scalasca/Score-P instrumentation on real platform executing benchmarks. This summary shows the degree of parallelism of the chosen workloads. Instruction summary shown in Figure \ref{fig:summary} (b) gives the percentage of different instruction executed over workload runtime, such as IntAlu, SimdFloat, MemRead and MemWrite instructions. 
It has been obtained through a cycle-approximate simulation with gem5. Figure \ref{fig:cpi} shows Cycles Per Instruction (CPI) per each workload executed on three cores, i.e. A7, A9 and A15.  

\section{Architecture exploration}
\label{sec:results}
This section is devoted to the presentation and analysis of the obtained simulation results for the chosen HSA configurations, according to two chosen metrics: execution time (referred to as \textit{Delay} in the following) and Energy-to-Solution.

\subsection{Results}
\begin{figure*}[!ht]
\centering
\subfloat[\textit{heartwall}]{\includegraphics[scale=0.5]{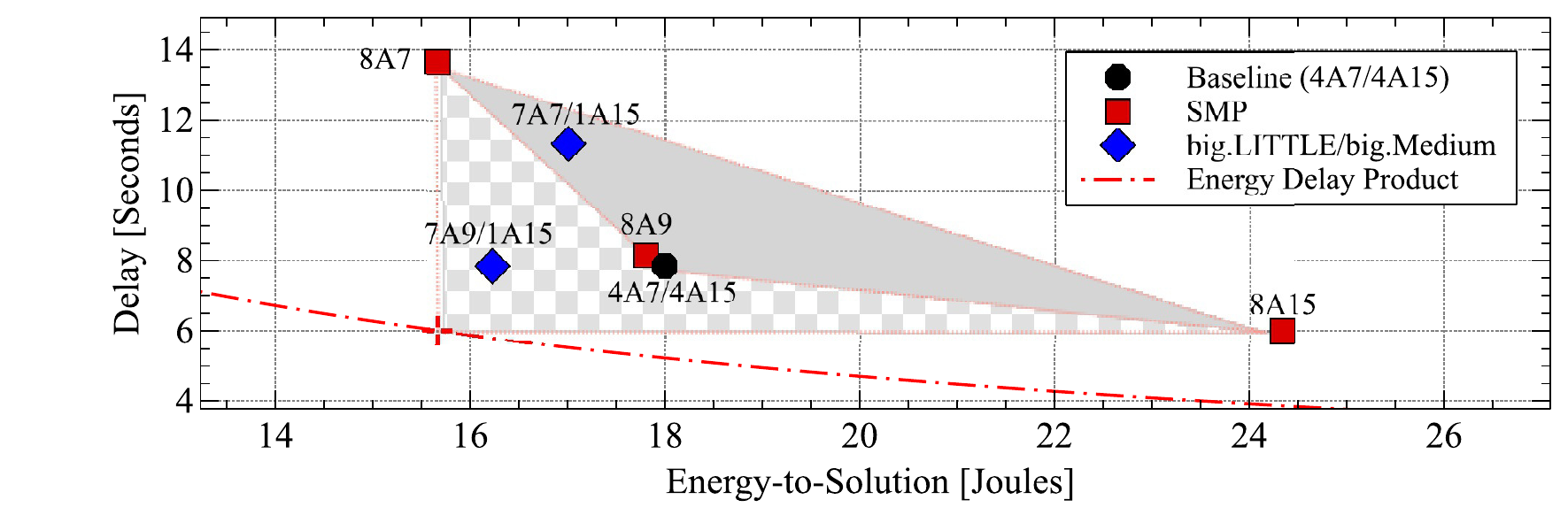}}
\subfloat[\textit{heartwall}]{\includegraphics[scale=0.5]{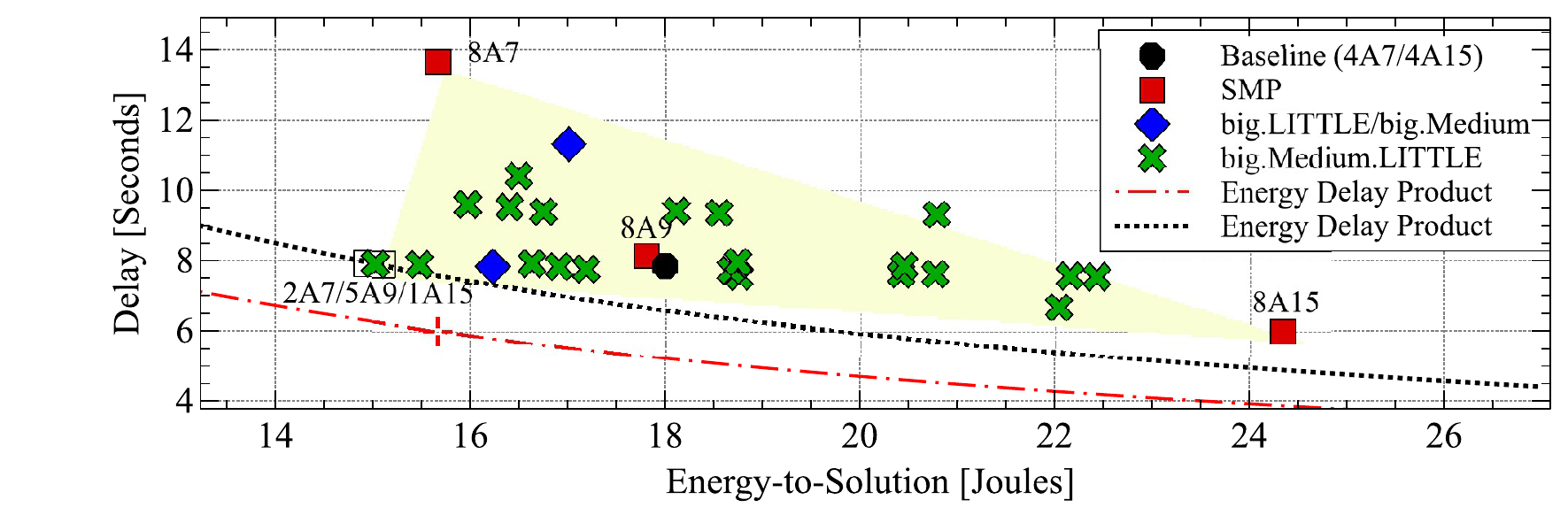}}
\\ \vspace{-2ex}
\subfloat[\textit{lud}]{\includegraphics[scale=0.5]{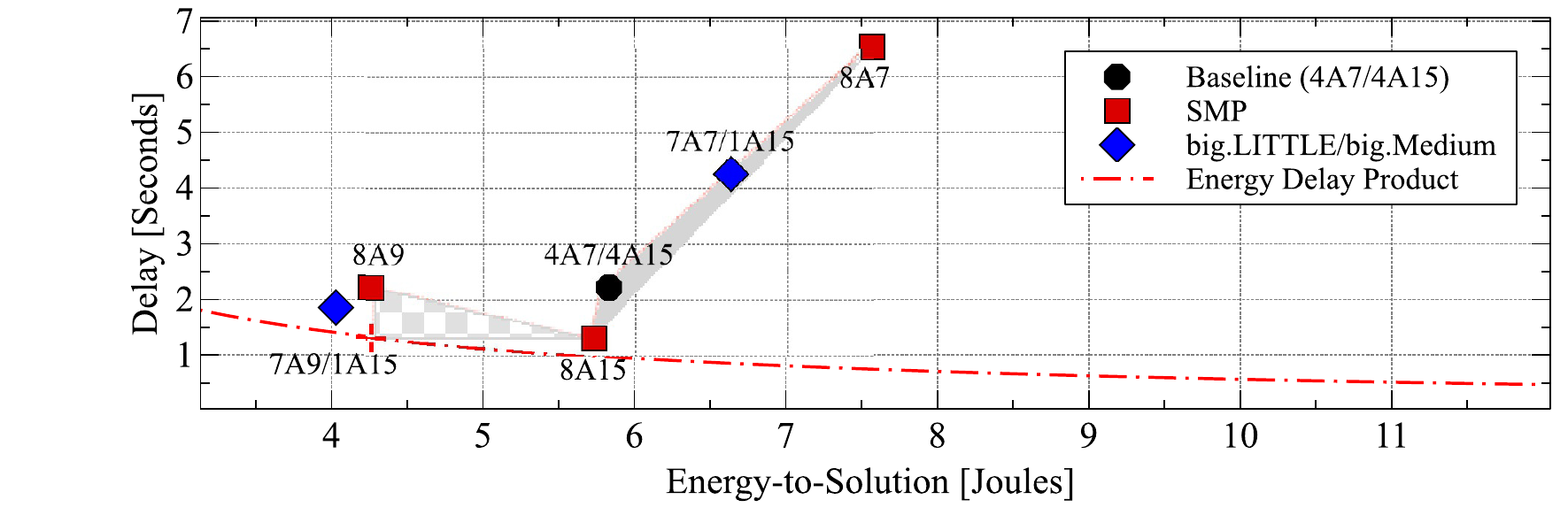}}
\subfloat[\textit{lud}]{\includegraphics[scale=0.5]{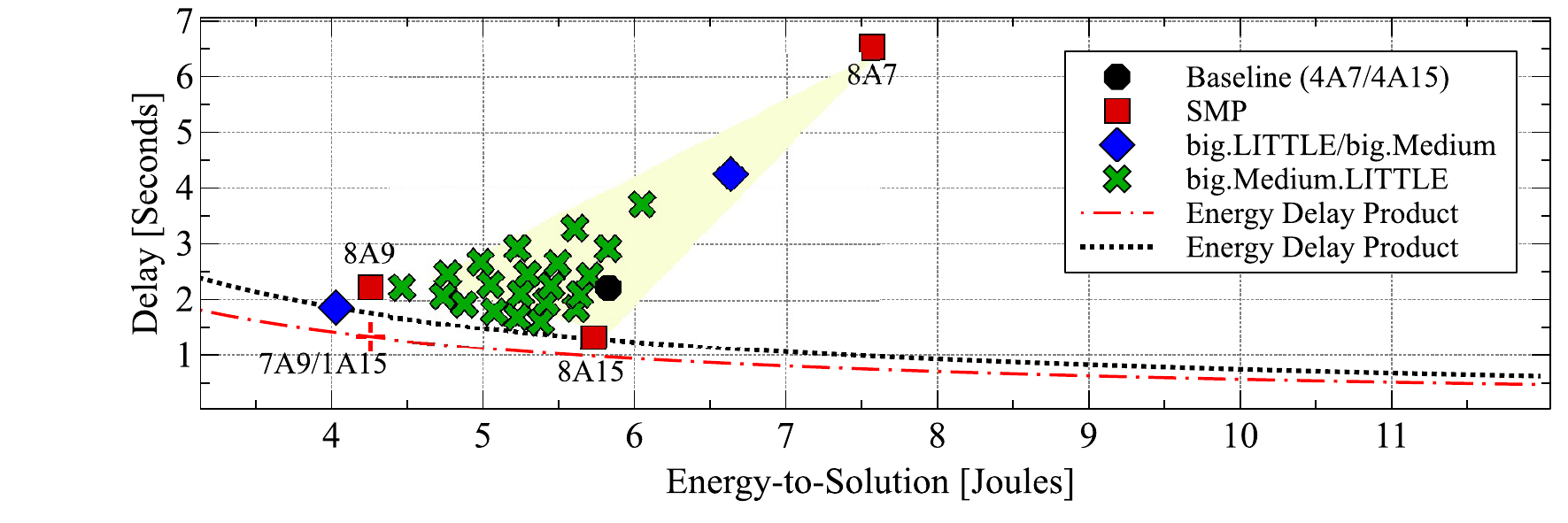}}
\\ \vspace{-2ex}
\subfloat[\textit{nw}]{\includegraphics[scale=0.5]{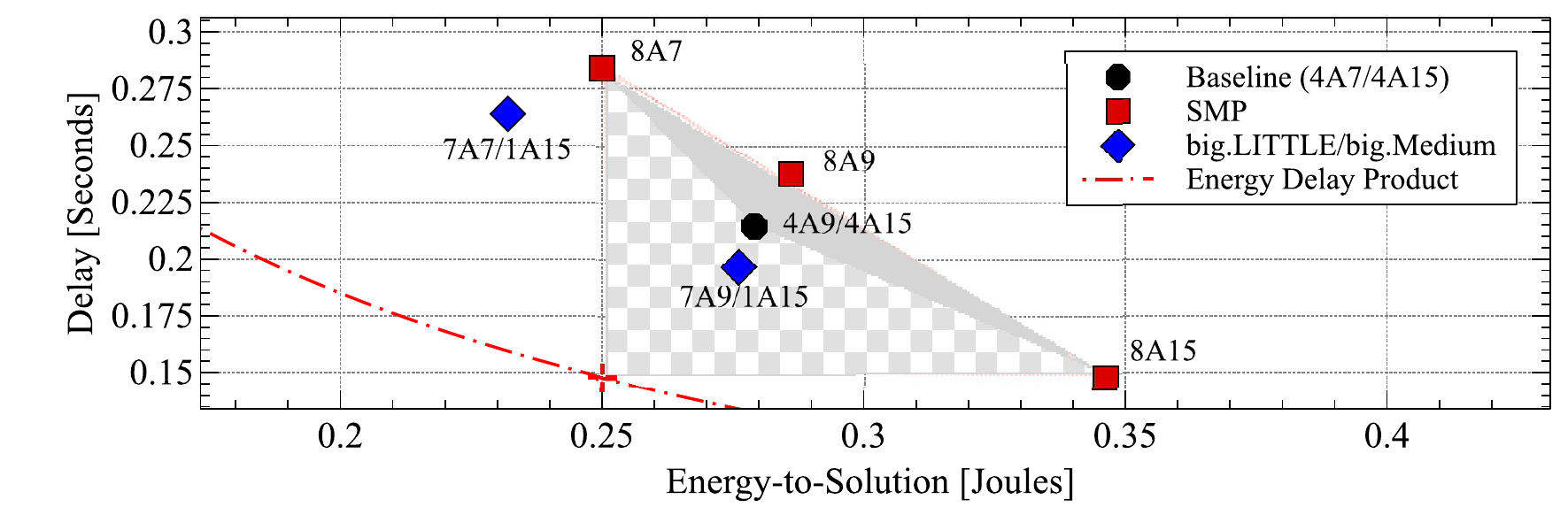}}
\subfloat[\textit{nw}]{\includegraphics[scale=0.5]{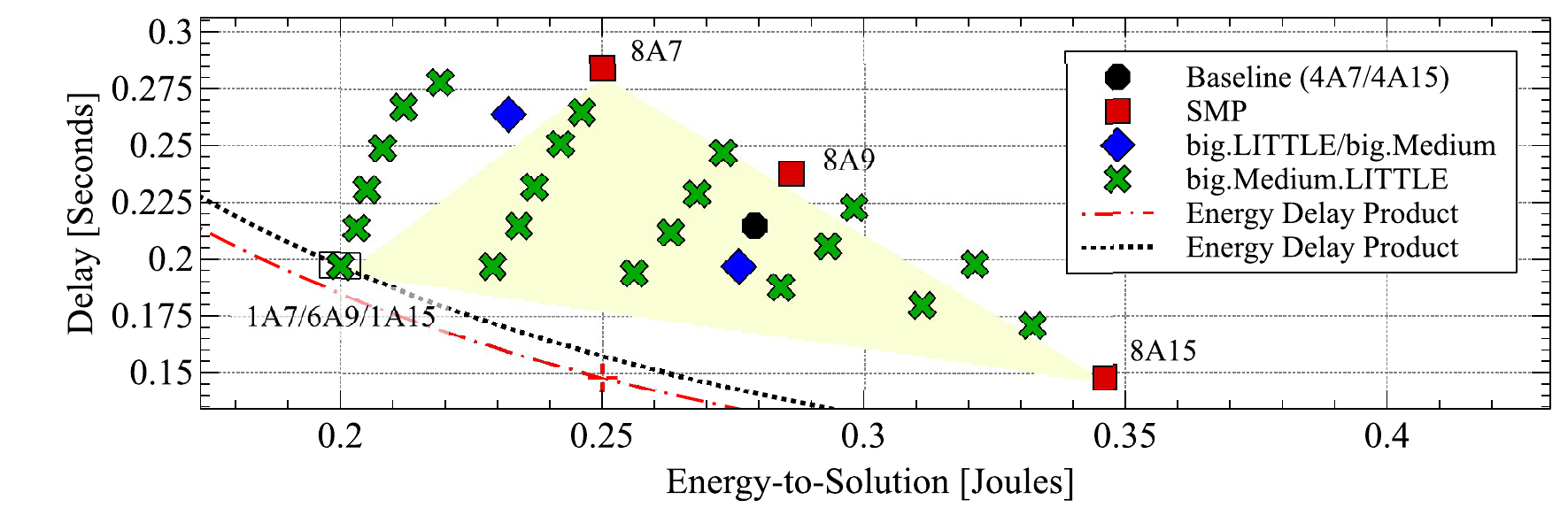}}
\\ \vspace{-2ex}
\subfloat[\textit{kmeans}]{\includegraphics[scale=0.5]{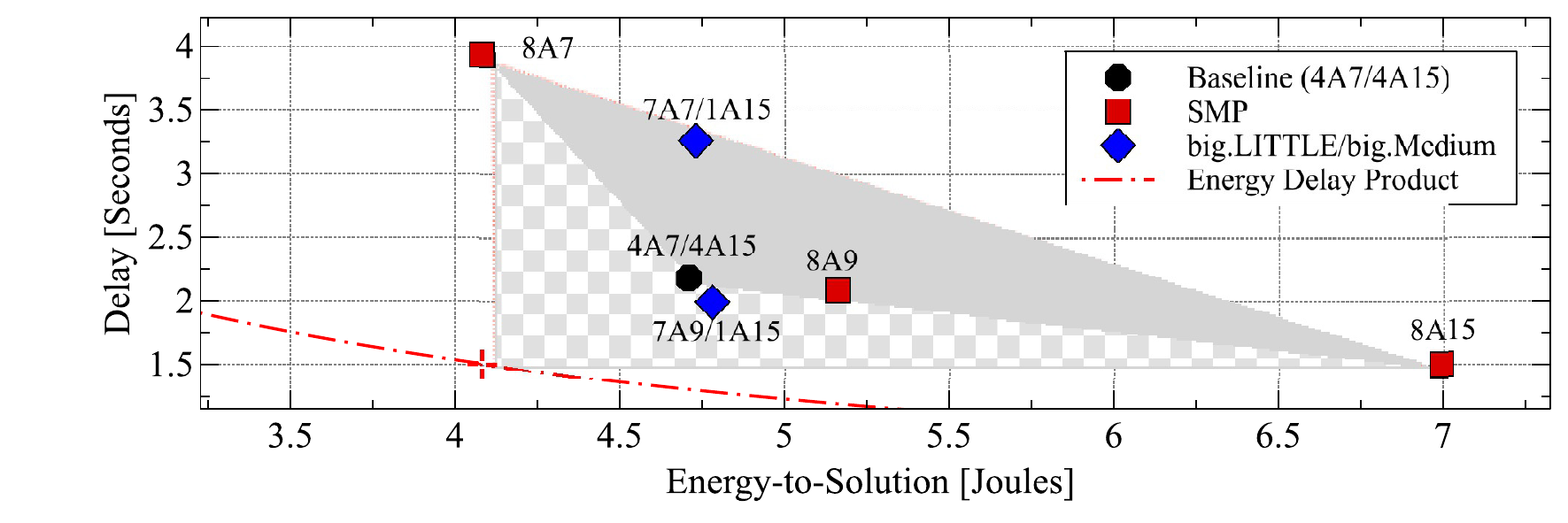}}
\subfloat[\textit{kmeans}]{\includegraphics[scale=0.5]{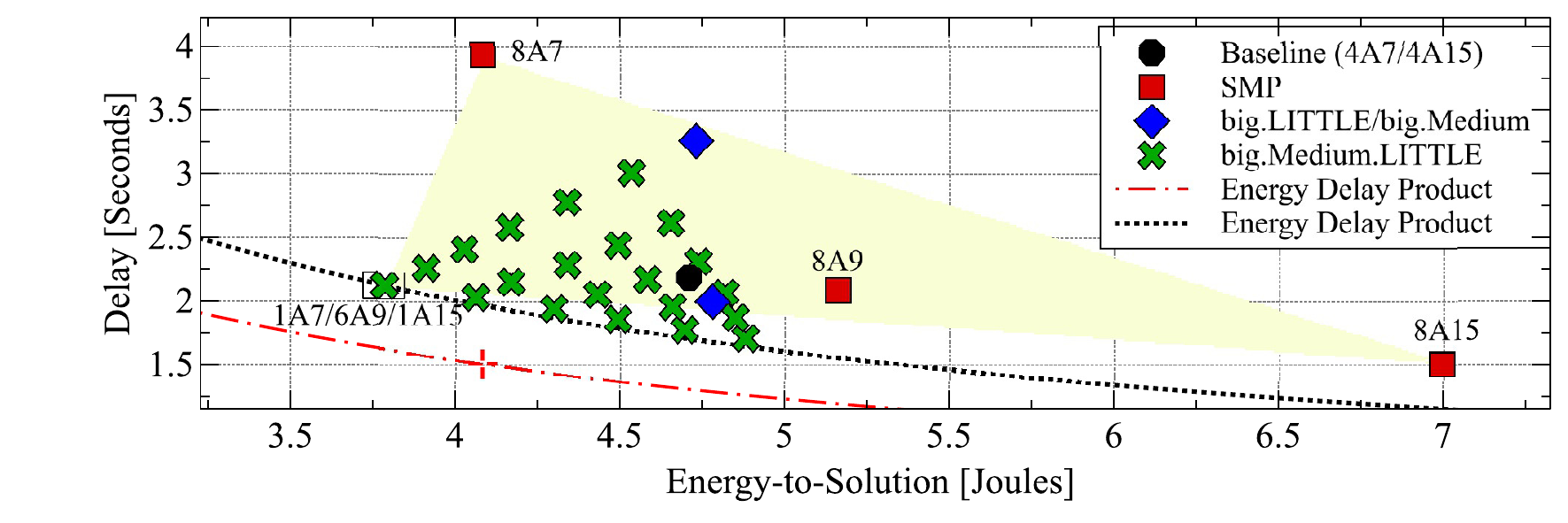}}
\caption{Performance and energy trade-offs (Part 1).}
\label{fig:HMP1}
\end{figure*}

\begin{figure*}[!ht]
\centering
\subfloat[\textit{nn}]{\includegraphics[scale=0.5]{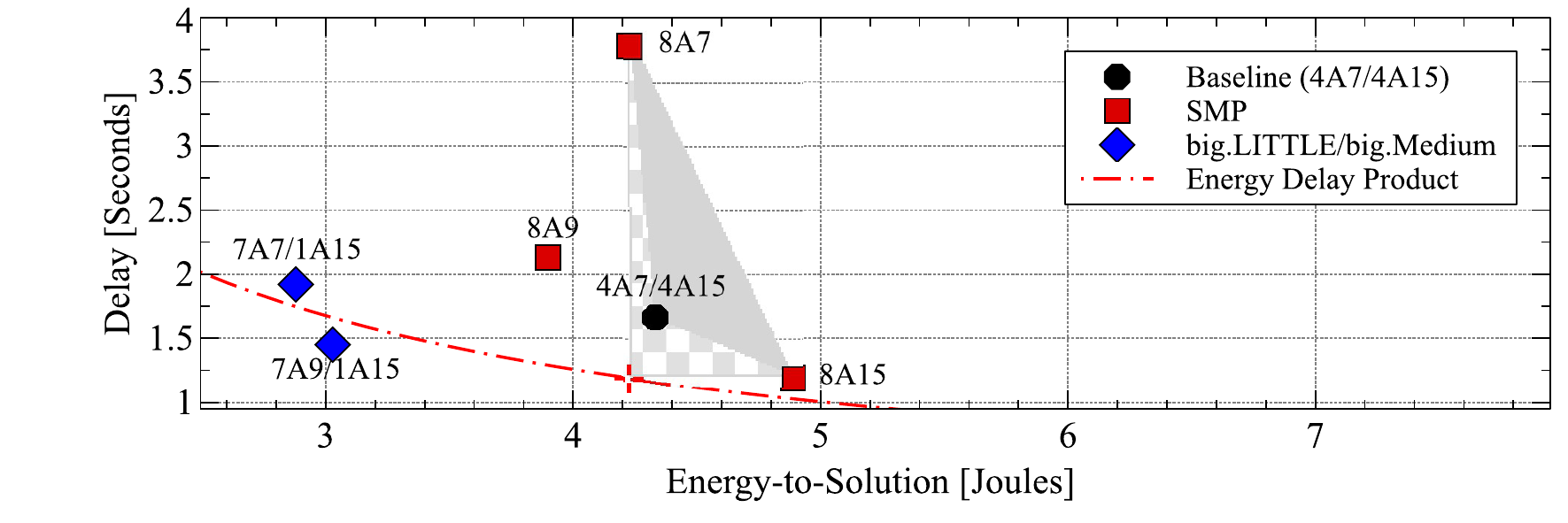}}
\subfloat[\textit{nn}]{\includegraphics[scale=0.5]{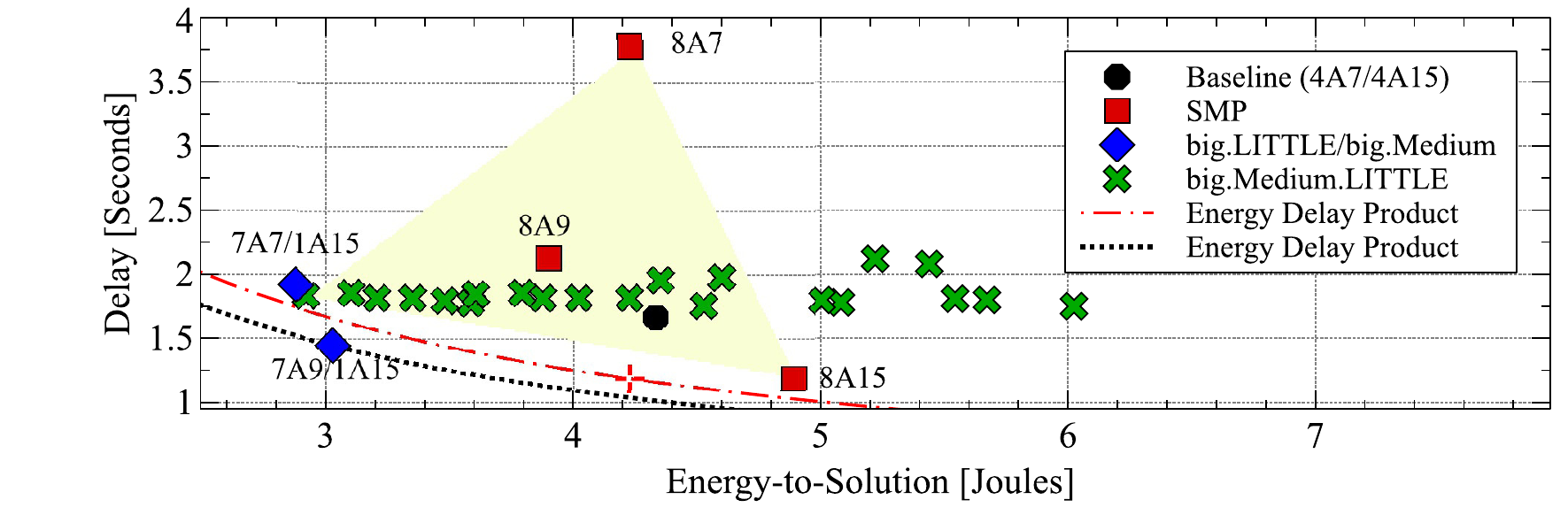}}
\\ \vspace{-2ex}
\subfloat[\textit{backprop}]{\includegraphics[scale=0.5]{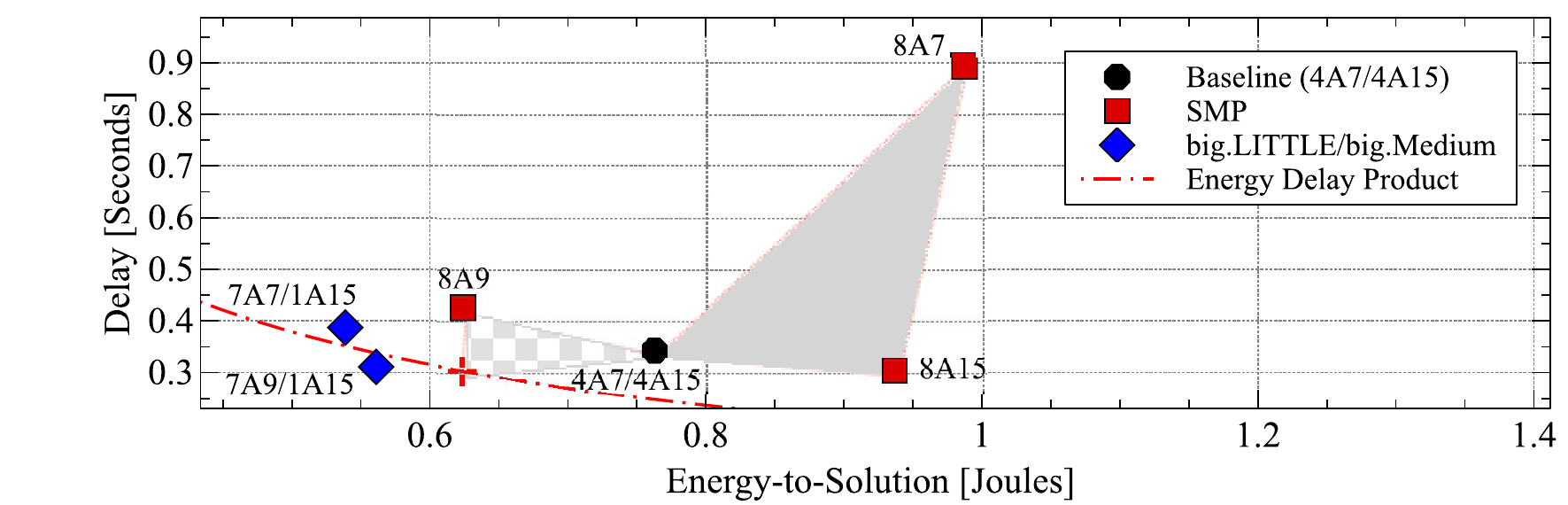}}
\subfloat[\textit{backprop}]{\includegraphics[scale=0.5]{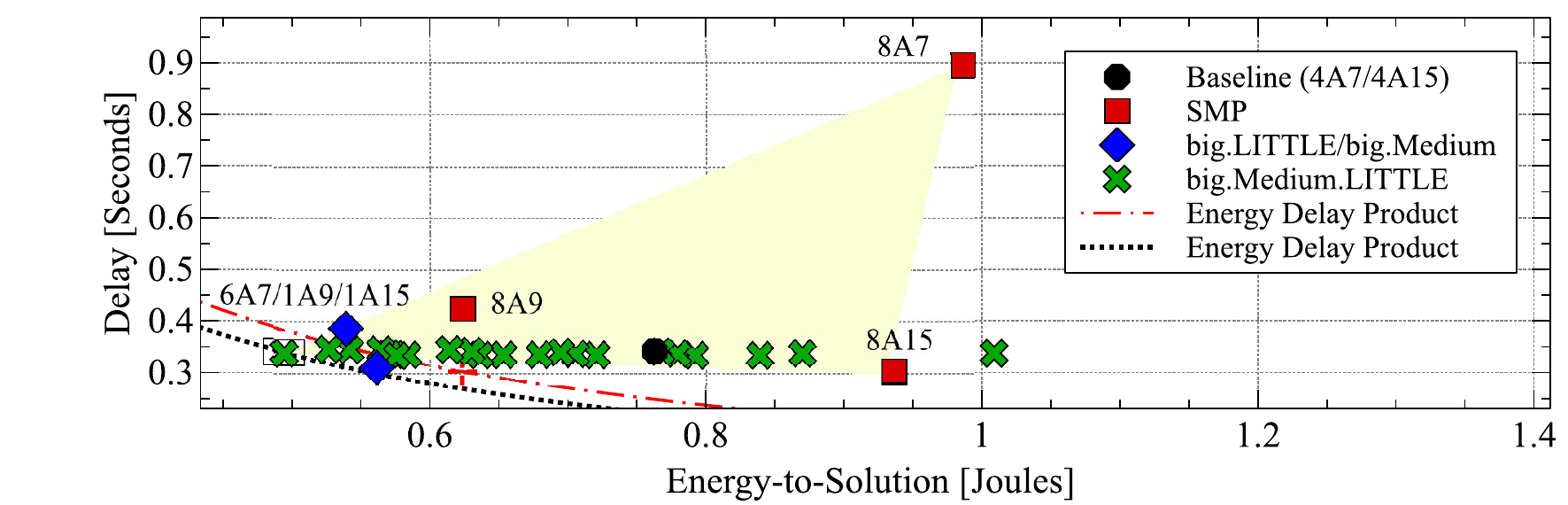}}
\\ \vspace{-2ex}
\subfloat[\textit{srad v1}]{\includegraphics[scale=0.5]{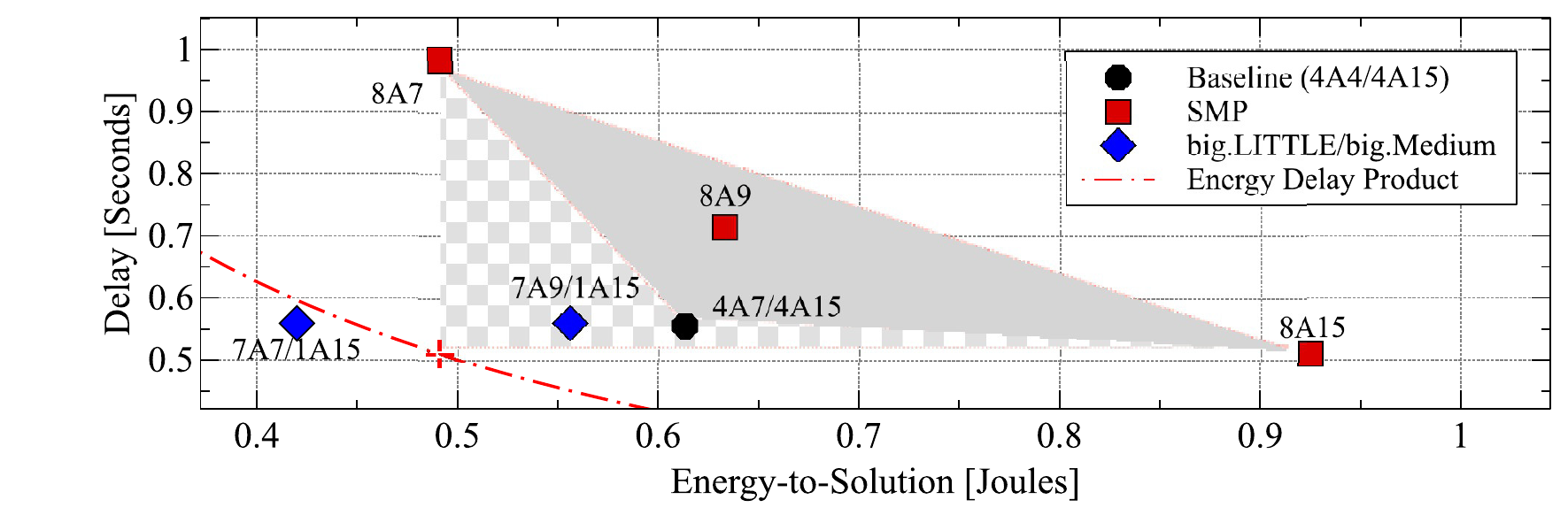}}
\subfloat[\textit{srad v1}]{\includegraphics[scale=0.5]{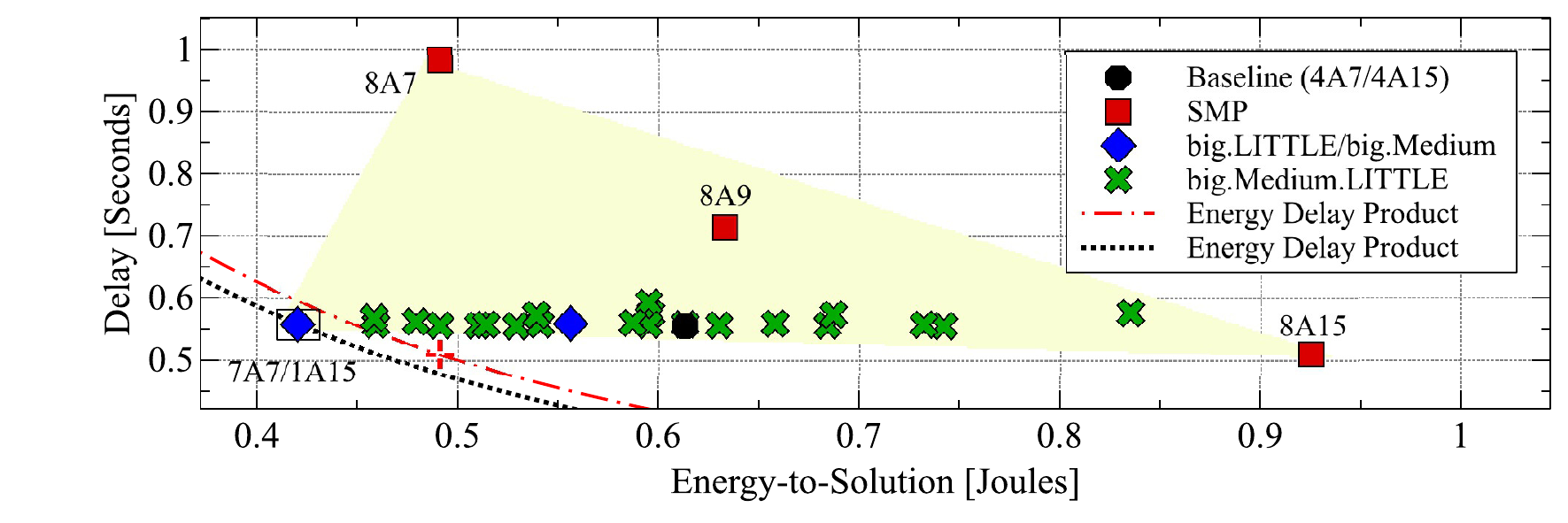}}
\caption{Performance and energy trade-offs (Part 2).}
\label{fig:HMP2}
\end{figure*}

Figures \ref{fig:HMP1} and \ref{fig:HMP2} present the trade-offs between the delay and EtoS for chosen \texttt{Rodinia} workloads running in gem5/McPAT simulation environment. 

Red dash-dotted line shows the EDP provided by the Ideal point as in Figure \ref{fig:projection}.  Black dotted line represents the points that achieve the best EDP according to the simulation results.

We analyze these results in three steps classifying the workloads according to the points placement. We aim to link the execution behavior on the top of SMP configurations with the $H^2$ and $H^3$ scenarios. 

\subsubsection{SMP points placement.} \label{subsec:511} We distinguish two categories of SMP points placement. The first category is characterized by the compromise between the computational performance and consumed energy that places A7, A9 and A15\footnote{In this section, the abbreviated notations are used for core types, i.e. A7 for Cortex-A7, A9 for Cortex-A9 and A15 for Cortex-A15.} points diagonally left-to-right, top-to-bottom. Such an arrangement is intuitively predictable and corresponds to the projection shown in Figure \ref{fig:projection}. Most of the presented workloads provide such a behavior, i.e. \textit{heartwall}, \textit{kmeans}, \textit{nw} and \textit{srad v1}.

Workloads that belong to the second category, such as \textit{backprop} and \textit{lud} have the 8A7 point placed in the upper right corner. That makes the 8A7 configuration the worse scenario in both metrics, i.e. performance and energy. The 8A9 point in turn constitutes the most energy-efficient configuration. In such scenarios we plot the Ideal point based on the A9/A15 couple instead of the A7/A15 couple. \textit{nn} application points has close arrangement. The 8A7 point shows the worse performance, but in terms of energy it outperforms the 8A15 configuration.

Presented workloads behavior strongly correlates with the average CPI evaluation shown in Figure \ref{fig:cpi}.  
\textit{lud} workloads obtains the most critical performance loss when using the A7 core, e.g. around 4x-4.5x CPI slowdown. The use of A9 core shows less variation among considered workloads providing 1.5x-2x slowdown.
The reason lies in the workloads computation nature and level of sensitivity to core micro-architecture; especially when switching from in-order A7 cores to out-of-order A15s. 

The \textit{lud} application belongs to dense linear algebra and is compute-bound. Moreover, it contains a rather important number of \textit{SimdFloat} instructions compared to other workloads. All these features lead to important penalty when switching to A7, but not to A9. This observation indicates that the reduced SIMD-width of 64 in A9 instead of 128 bits in A15 does not have a big impact on performance. \textit{kmeans} and \textit{nn} while being also compute-bound demonstrate moderate slowdown when switching from A15 to A7 or to A9. Compared to \textit{lud}, they have 20\% of sequential execution and 10-20\% of time spent in synchronization (see Figure \ref{fig:summary} (a)). \textit{backprop} and \textit{heartwall} strongly depend on the memory infrastructure (see Table \ref{tab:problem_size}). In \textit{backprop} application, this dependence combined with a predominant serial region leads to the worst performance for A7 than for A9 and A15.\label{par1}  
\textit{nw} and \textit{srad v1} are also memory latency/bandwidth limited and show the least important slowdown around 2x. This value corresponds to the theoretical performance ratio shown in Figure \ref{fig:ratio}. 

In addition to CPI slowdown, we compare the gain in terms of power consumption. We observe that several workloads benefit more being run in A7 and others benefit less. Dense linear algebra workloads, such as \textit{kmeans}, \textit{lud}, and \textit{nn} are compute-intensive and result in higher power consumption and temperature. Thus, the gain while switching from A15 cores to A7 cores is more important. Memory latency-bound workloads, i.e. \textit{backprop} and \textit{nw}, both make extensive use of shared memory and compared to the previous group of compute-intensive workloads show lower gain. Memory bandwidth limited workloads, i.e. \textit{heartwall} and \textit{srad v1}, exhibit massive parallelism. According to the collected statistics, the memory bandwidth constrain has not been reached over the simulations. Thereby, the power-efficiency of A7 configuration is relatively important. 

\subsubsection{big.LITTLE points placement.} Three configurations, i.e. symmetric big.LITTLE baseline (4A7/4A15), asymmetric big.LITTLE (7A7/1A15) and asymmetric big.Medium (7A9/1A15) are considered here.

The baseline configuration is expected to provide a performance/energy trade-off placing the point in a checkerboard triangle. We observe this result in most cases, i.e. \textit{heartwall}, \textit{kmeans}, \textit{nn}, \textit{nw}, and \textit{srad v1}. Workload that shows untypical SMP points placement, i.e. \textit{backprop}, also benefit from $H^2$ configurations although significant inefficiency of SMP 8A7. According to the functional summary presented in Figure \ref{fig:summary} a), this workload possesses a low degree of parallelism - the main thread takes 70\% of execution time. Thus at least one big core is required for efficient workload execution. One more workload with untypical SMP points placement, e.g. \textit{lud}, presents an high degree of parallelism and does not provide suitable performance/energy trade-off being executed in 4A7/4A15 configuration as well as on asymmetric 7A7/1A15. However, it benefits from A9/A15 combination.

In most cases, asymmetric configurations provide suitable trade-offs in performance and energy. Especially, workloads with low degree of parallelism obtain such a good balance in performance and energy that resulting EDP gets very close to the Ideal scenario and event outperforms it, e.g. \textit{srad v1}.

Workloads with important portion of \textit{SimdFloat} operations shown in Figure \ref{fig:summary} b), such as  \textit{heartwall}, \textit{kmeans} and \textit{lud}, lose efficiency when increasing the number of A7 cores. However, combining A15 with A9 we observe significant enhancement in performance and energy trade-off. \textit{lud} application shows the best example possessing the highest percentage of \textit{SimdFloat} operations among the workloads mentioned above. 

\subsubsection{big.Medium.LITTLE points placement.} The right column in Figure \ref{fig:HMP1} and Figure \ref{fig:HMP2} show the placement of $H^3$ configurations points. 
The yellow triangular joins two basic SMP points, i.e. 8A7 and 8A15, with the big.Medium.LITTLE point that provides the best EDP result among other $H^3$ points. The black dotted line shows the iso-EDP curve corresponding to the best achieved result among all considered configurations, e.g. SMP, $H^2$ or $H^3$.

Executing \textit{lud} workload, all configurations that contain the A7 core show to be inefficient due to a lot of \textit{SimdFloat} operations resulting in a high CPI. $H^3$ points are placed within the yellow triangular borders. The trade-off between the performance and energy correlates with the number of A7, A9 and A15 cores. Two configurations provide the best EDP result, i.e. SMP A15 and $H^2$ 7A9/1A15. 

\textit{nn} application is compute-bound and benefits more from being executed on A9 cores rather than on A7. It spends 25\% of time in synchronization due to the control flow divergence. \textit{nn} shows to be less efficient while running on $H^3$ configurations. Higher number of A15 cores does not improve the performance resulting in higher power consumption and no performance gain. The best EDP is provided by 7A9/1A15 configuration. This outlier outperforms the baseline by 38\% and the ideal scenario by 14\%. 

\textit{kmeans} and \textit{nw} provide trade-off points placed in a clearly outlined triangular. Each row of points corresponds to configurations with a fixed number of one type of cores. These workloads have several common features such as high degree of parallelism, low synchronization overhead, typical SMP points placement. In both cases, the point placed at the left vertex of the formed triangular corresponds to the configuration 1A7/6A9/1A15. 
This configuration provides the best EDP result and outperforms the existing baseline, i.e. 4A7/4A15, by 23\% and 34\% respectively.
The 7A9/1A15 configuration that provide the best EDP for previously discussed workloads is also outperformed. This can be explained by the bandwidth limitation which does not allow to get the best from this configuration.

\textit{heartwall} workload also has these common characteristics. The application does not form a clearly outlined shape. 
Even if the 1A7/6A9/1A15 configuration is better than the baseline, the best EDP result is provided by 2A7/5A9/1A15 configuration that outperforms the baseline by 17\%. The bandwidth limitation implies to replace 2A9 cores by 2A7 in this case. 

Two workloads with the lowest degrees of parallelism, e.g. \textit{backprop} and \textit{srad v1}, provide better EDP while running on the $H^2$ 7A7/1A15 or $H^3$ 6A7/1A9/1A15. Cores with potentially better performance, e.g. A9 or A15, do not improve delay much but incur a notable energy penalty. Configurations that provide the best EDP results outperform the baseline by around 30\% and the theoretical ideal scenario by 8\%.

\subsection{Observations and learnings}

We summarize the main observations from the previous exploration and give some insights.

\subsubsection{Main observations}
\label{subsubsec:521}
\textbf{General observation:} Varying the number of big and LITTLE cores as well as extending heterogeneity with the third Medium cluster, we demonstrated that with alternative configurations significant improvement in performance and energy can be achieved, up to 38\% against the baseline. Among considered SMP and HMP scenarios, we distinguished several alternative configurations, which outperform the baseline running different workloads, e.g. 7A7/1A15, 7A9/1A15, 1A7/6A9/1A15, 2A7/5A9/1A15, 6A7/1A9/1A15, 8A15. $H^3$ configurations achieve better energy/delay trade-offs in most cases. 

\textbf{Impact of serial regions:}
For all workloads with predominant serial regions e.g. \textit{backprop}, \textit{nn}, and \textit{srad v1}, all configurations with at least one A15 core provide very close delay values, e.g. for \textit{srad v1} workload the delay is around 0.55s (see Figure \ref{fig:HMP2} (e) and (f)). This is due to the fact that the A15 is used to execute these regions. 
Therefore, it is advisable to include one big core cluster on each configuration.

\textbf{Impact of parallel regions:}
All workloads with preponderant parallel regions have a triangle-shape distribution e.g. \textit{kmeans}, \textit{lud}, and \textit{nw}. The best configuration depends among other parameters on the nature of the application.

\textbf{Impact of computation complexity:}
For computationally intensive workloads e.g. \textit{lud} and \textit{nn}, in-order cores present high CPI which degrades performance. They are therefore uninteresting for such types of workloads. 7A9/1A15 configuration proposes best EtoS and EDP which will favor this configuration even if 8A15 is slightly better in term of delay. The high number of SimdFloat operations in \textit{lud} confirms this since the A7 cluster is the least efficient in terms of SIMD-performance/power of the considered core types. 

\textbf{Impact of communication intensity:}
Regarding workloads limited by the memory capabilities, a balance is established between memory latency and bandwidth. Indeed, if the workload is only limited by latency, out-of-order cores such as A9 provide better energy-efficiency than A7 e.g. \textit{nn} workload.
For  memory bandwidth-limited workloads, higher numbers of A7 cores are more beneficial due to the lower power consumption of this core type. Note that this observation relate to the chosen memory subsystem only (which is same for all configurations) and would likely differ much for systems with many more cores, or faster memory subsystem.

\textbf{Impact of core configurations on chip area:} Using the ratios shown in Figure \ref{fig:ratio}, we estimate the impact of alternative configurations on the chip area. The gain in terms of silicon for 7A7/1A15, 7A9/1A15, 1A7/6A9/1A15, 2A7/5A9/1A15, 6A7/1A9/1A15 configurations versus 4A15/4A7 baseline are 2.27x, 1.33x, 1.49x, 1.58 and 2.1x respectively.

\subsubsection{Gained insights}
The following insights can be drawn:
\begin{compactitem}
\item Each configuration must include one big core to process the serial regions of the workloads.

\item For compute intensive workloads, out-of-order cores are preferred especially if the workloads include floating-point operations. Indeed, A7 cores feature lower arithmetic performance compared to A9 and A15.

\item For memory latency limited workloads, A9 cores are the most suitable. The A15s are more power-hungry than the A9s thereby incurring a significant energy overhead. And the A7 cores do not benefit of the dynamic execution advantage of the out-of-order cores that can help to hide memory latency. 
											
\item For memory bandwidth limited workloads, A7 cores are the best candidates. Their power consumption is lower than other core types and since cores spend significant time idle, waiting for data due to bandwidth limitations, the use of faster (and more power-hungry) cores does not bring any advantage.

\item Since a workload is generally composed of several regions, each of which may exhibit different characteristics (computation bound, latency bound or bandwidth-limited), best configurations include for most applications three cores types.
\end{compactitem}

From the above summary, we notably observe that configurations with 3 core types often outperform those with 2 types.
Further, some of the conclusions drawn above are subject to the memory subsystem performance. This applies notably for memory bandwidth-limited workloads in which A7s are preferred for their low-power consumption and negligible penalty on performance. One interesting direction lies in deciding the HSA configuration taking into account the existing memory subsystem as described in \cite{6531079} for time predictability, so as to pick cores that provide best compute performance yet without saturating the memory bandwidth.

Even though the above insights were made on the basis of static HSA configurations executing conventional workloads, these reveal the potential of matching benchmark nature to hardware configuration and suggest perspectives for dynamic architecture reconfiguration. 
From an application perspective though, some configurations would not be acceptable because they do not meet the level of performance required (i.e. execution time). Quality-tunable applications / algorithms circumvent this limitation by accepting a lower output quality resulting in lesser compute complexity. Not only would this allow picking some very low power configurations identified above (yet matching the performance requirements), but this would altogether permit to activate only a subset of the available cores, such as 2 A7 and 1 A9 out of the 8 available.

In video processing domain for instance, sub-optimal algorithms for motion estimation (e.g., SLIMPEG motion estimation \cite{Alfonso2002UltraLM}) sacrifice the precision (or quality) of results for mitigating computation complexity (and reducing  both execution time overhead and power consumption), could be preferably processed by selecting configurations with little and medium cores. Optimal algorithms (Full Search motion estimation algorithms \cite{Monteiro:2014:PFS:2589752.2589763}), which usually involve complex computations for determining more precise results would require configurations combining big cores with other core types.


\section{Conclusions and Future Work}
%
%
\label{sec:conclusions}
In this paper, we studied the impact of architecture heterogeneity in single-ISA HSAs on the performance and energy for a set of scientific workloads. As a baseline, we used previously validated model of ARM big.LITTLE platform implemented in gem5 and McPAT simulation frameworks. Experimental results demonstrated that varying the level of architecture heterogeneity result in significant performance and energy improvements, i.e. up to 2.3x and 1.5x respectively against the baseline. Based on the workload profile, we provided useful insight on how application characteristics (level of parallelism, computation complexity, communication pattern) determine performance/energy trade-off and what are the most suitable architecture configurations for these different workloads. This contributes to understanding the application requirements impact and can guide future research towards dynamically reconfigurable HSAs. 
The insights gained from the present study can be leveraged in energy-efficient designs for executing quality-tunable algorithms, by carefully selecting a rather small subset of cores, i.e. those which provide the best energy efficiency for instance.


%

\section*{Acknowledgment}

The research leading to these results has received funding from the European Community's H2020 Program under the Mont-Blanc 3 Project (www.montblanc-project.eu), grant agreement n\textsuperscript{o} 671697.

\ifCLASSOPTIONcaptionsoff
  \newpage
\fi



%
{\footnotesize
\bibliographystyle{IEEEtran}
\bibliography{IEEEabrv,paper} 

\begin{thebibliography}{10}
\providecommand{\url}[1]{#1}
\csname url@samestyle\endcsname
\providecommand{\newblock}{\relax}
\providecommand{\bibinfo}[2]{#2}
\providecommand{\BIBentrySTDinterwordspacing}{\spaceskip=0pt\relax}
\providecommand{\BIBentryALTinterwordstretchfactor}{4}
\providecommand{\BIBentryALTinterwordspacing}{\spaceskip=\fontdimen2\font plus
\BIBentryALTinterwordstretchfactor\fontdimen3\font minus
  \fontdimen4\font\relax}
\providecommand{\BIBforeignlanguage}[2]{{%
\expandafter\ifx\csname l@#1\endcsname\relax
\typeout{** WARNING: IEEEtran.bst: No hyphenation pattern has been}%
\typeout{** loaded for the language `#1'. Using the pattern for}%
\typeout{** the default language instead.}%
\else
\language=\csname l@#1\endcsname
\fi
#2}}
\providecommand{\BIBdecl}{\relax}
\BIBdecl

\bibitem{Monteiro:2014:PFS:2589752.2589763}
\BIBentryALTinterwordspacing
E.~Monteiro, B.~Vizzotto, C.~Diniz, M.~Maule, B.~Zatt, and S.~Bampi,
  ``Parallelization of full search motion estimation algorithm for parallel and
  distributed platforms,'' \emph{Int. J. Parallel Program.}, vol.~42, no.~2,
  pp. 239--264, Apr. 2014. [Online]. Available:
  \url{http://dx.doi.org/10.1007/s10766-012-0216-7}
\BIBentrySTDinterwordspacing

\bibitem{Alfonso2002UltraLM}
D.~Alfonso, A.~Artieri, A.~Capra, M.~E. Mancuso, F.~Pappalardo, F.~S. Rovati,
  and R.~Zafalon, ``Ultra low-power multimedia processor for mobile multimedia
  applications,'' \emph{Proceedings of the 28th European Solid-State Circuits
  Conference}, pp. 63--69, 2002.

\bibitem{7368023}
J.~M. Shalf and R.~Leland, ``Computing beyond {Moore's Law},'' \emph{Computer},
  vol.~48, no.~12, pp. 14--23, Dec 2015.

\bibitem{DBLP:journals/corr/Lavin15b}
A.~Lavin and S.~Gray, ``Fast algorithms for convolutional neural networks,'' in
  \emph{Proceedings of the IEEE Conference on Computer Vision and Pattern
  Recognition}, 2016, pp. 4013--4021.

\bibitem{Jouppi:2017:IPA:3079856.3080246}
N.~P. Jouppi, C.~Young, N.~Patil, D.~Patterson, G.~Agrawal, R.~Bajwa, S.~Bates,
  S.~Bhatia, N.~Boden, A.~Borchers \emph{et~al.}, ``In-datacenter performance
  analysis of a tensor processing unit,'' in \emph{Proceedings of the
  International Symposium on Computer Architecture}, 2017, pp. 1--12.

\bibitem{DBLP:conf/ets/HanO13}
J.~Han and M.~Orshansky, ``Approximate computing: An emerging paradigm for
  energy-efficient design,'' in \emph{{ETS}}.\hskip 1em plus 0.5em minus
  0.4em\relax {IEEE} Computer Society, 2013, pp. 1--6.

\bibitem{Ueng2008}
S.-Z. Ueng, M.~Lathara, S.~S. Baghsorkhi, and W.~H. Wen-mei, ``{CUDA-lite:
  Reducing GPU programming complexity},'' in \emph{Proceedings of teh
  International Workshop on Languages and Compilers for Parallel Computing},
  2008, pp. 1--15.

\bibitem{Kumar}
R.~Kumar, D.~M. Tullsen, P.~Ranganathan, N.~P. Jouppi, and K.~I. Farkas,
  ``Single-{ISA} heterogeneous multi-core architectures for multithreaded
  workload performance,'' in \emph{Proceedings of the International Symposium
  on Computer Architecture}, 2004, pp. 64--75.

\bibitem{Samsung:Exynos}
Samsung, ``{Exynos Octa SoC},'' https://http://www.samsung.com/, [Accessed:
  May-2018].

\bibitem{qualcomm}
{Qualcomm Technologies, Inc.}, ``Qualcomm snapdragon,''
  https://www.qualcomm.com/products/snapdragon/, [Accessed: May-2018].

\bibitem{Nvidia:Tegra}
Nvidia, ``Tegra mobile processors,'' http://www.nvidia.com, [Accessed:
  May-2018].

\bibitem{ChronakiRBALV15}
K.~Chronaki, A.~Rico, R.~M. Badia, E.~Ayguad{\'e}, J.~Labarta, and M.~Valero,
  ``Criticality-aware dynamic task scheduling for heterogeneous
  architectures,'' in \emph{Proceedings of the International Conference on
  Supercomputing}, 2015, pp. 329--338.

\bibitem{openmp}
{OpenMP Architecture Review Board}, ``The {OpenMP API} specification for
  parallel programming,'' http://openmp.org, [Accessed: May-2018].

\bibitem{VanCraeynest:2013:UFD:2400682.2400691}
K.~Van~Craeynest and L.~Eeckhout, ``Understanding fundamental design choices in
  single-{ISA} heterogeneous multicore architectures,'' \emph{ACM Transactions
  on Architecture and Code Optimization}, vol.~9, no.~4, p.~32, 2013.

\bibitem{Takouna_efficientvirtual}
I.~Takouna, W.~Dawoud, and C.~Meinel, ``Efficient virtual machine
  scheduling-policy for virtualized heterogeneous multicore systems,'' in
  \emph{Proceedings of the International Conference on Parallel and Distributed
  Processing Techniques and Applications}, 2011.

\bibitem{Koufaty:2010:BSH:1755913.1755928}
D.~Koufaty, D.~Reddy, and S.~Hahn, ``Bias scheduling in heterogeneous
  multi-core architectures,'' in \emph{Proceedings of the European conference
  on Computer systems}, 2010, pp. 125--138.

\bibitem{Binkert:2011:GS:2024716.2024718}
N.~Binkert, B.~Beckmann, G.~Black, S.~K. Reinhardt, A.~Saidi, A.~Basu,
  J.~Hestness, D.~R. Hower, T.~Krishna, S.~Sardashti \emph{et~al.}, ``The gem5
  simulator,'' \emph{ACM SIGARCH Computer Architecture News}, vol.~39, no.~2,
  pp. 1--7, 2011.

\bibitem{McPAT}
HP, ``{McPAT},'' http://www.hpl.hp.com/research/mcpat/, [Accessed: May-2018].

\bibitem{mcsoc}
A.~Butko, F.~Bruguier, A.~Gamati{\'e}, G.~Sassatelli, D.~Novo, L.~Torres, and
  M.~Robert, ``Full-system simulation of {big.LITTLE} multicore architecture
  for performance and energy exploration,'' in \emph{Proceedings of the
  International Symposium on Embedded Multicore/Many-core Systems-on-Chip},
  2016, pp. 201--208.

\bibitem{5306797}
S.~Che, M.~Boyer, J.~Meng, D.~Tarjan, J.~Sheaffer, S.-H. Lee, and K.~Skadron,
  ``Rodinia: A benchmark suite for heterogeneous computing,'' in
  \emph{Proceedings of the International Symposium on Workload
  Characterization}, Oct 2009, pp. 44--54.

\bibitem{nvidiawp}
{Nvidia}, ``White paper: The benefits of quad core {CPUs} in mobile devices,''
  [Accessed: May-2018].

\bibitem{dal}
A.~Seznec, ``{Defying Amdahl’s Law (DAL)},''
  https://team.inria.fr/alf/members/andre-seznec/, [Accessed: May-2018].

\bibitem{armwp}
{Arm Ltd.}, ``White paper: big.{LITTLE} technology: The future of mobile,''
  https://www.arm.com/files/pdf/, [Accessed: May-2018].

\bibitem{Ipek:2007:CFA:1273440.1250686}
E.~Ipek, M.~Kirman, N.~Kirman, and J.~F. Martinez, ``Core fusion: accommodating
  software diversity in chip multiprocessors,'' \emph{ACM SIGARCH Computer
  Architecture News}, vol.~35, no.~2, pp. 186--197, 2007.

\bibitem{Kim:2007:CLP:1331699.1331733}
C.~Kim, S.~Sethumadhavan, M.~S. Govindan, N.~Ranganathan, D.~Gulati, D.~Burger,
  and S.~W. Keckler, ``Composable lightweight processors,'' in
  \emph{Proceedings of the International Symposium on Microarchitecture}, 2007,
  pp. 381--394.

\bibitem{Watanabe:2010:WWD:1816038.1815965}
Y.~Watanabe, J.~D. Davis, and D.~A. Wood, ``{WiDGET}: Wisconsin decoupled grid
  execution tiles,'' in \emph{ACM SIGARCH Computer Architecture News}, vol.~38,
  no.~3, 2010, pp. 2--13.

\bibitem{6493629}
M.~A. Suleman, M.~Hashemi, C.~Wilkerson, Y.~N. Patt \emph{et~al.}, ``Morphcore:
  An energy-efficient microarchitecture for high performance {ILP} and high
  throughput {TLP},'' in \emph{Proceedings of the International Symposium on
  Microarchitecture}, 2012, pp. 305--316.

\bibitem{Petrica:2013:FDA:2508148.2485924}
P.~Petrica, A.~M. Izraelevitz, D.~H. Albonesi, and C.~A. Shoemaker, ``Flicker:
  A dynamically adaptive architecture for power limited multicore systems,'' in
  \emph{ACM SIGARCH computer architecture news}, vol.~41, no.~3, 2013, pp.
  13--23.

\bibitem{Tavana:2015:EED:2744769.2744833}
M.~K. Tavana, M.~H. Hajkazemi, D.~Pathak, I.~Savidis, and H.~Homayoun,
  ``Elasticcore: enabling dynamic heterogeneity with joint core and
  voltage/frequency scaling,'' in \emph{Proceedings of the Design Automation
  Conference}, 2015, pp. 1--6.

\bibitem{arm_software}
\BIBentryALTinterwordspacing
{Arm Ltd.}, ``White paper: big.{LITTLE} technology moves towards fully
  heterogeneous global task scheduling.'' [Online]. Available:
  \url{http://www.arm.com/files/pdf/}
\BIBentrySTDinterwordspacing

\bibitem{6864009}
K.~Yu, D.~Han, C.~Youn, S.~Hwang, and J.~Lee, ``Power-aware task scheduling for
  {big.LITTLE} mobile processor,'' in \emph{Proceedings of the International
  SoC Design Conference}, 2013, pp. 208--212.

\bibitem{7059077}
C.~Tan, T.~S. Muthukaruppan, T.~Mitra, and L.~Ju, ``Approximation-aware
  scheduling on heterogeneous multi-core architectures,'' in \emph{Proceedings
  of the Asia and South Pacific Design Automation Conference}, 2015, pp.
  618--623.

\bibitem{butko2017efficient}
A.~Butko, F.~Bruguier, A.~Gamati{\'e}, and G.~Sassatelli, ``Efficient
  programming for multicore processor heterogeneity: Openmp versus ompss,'' in
  \emph{OpenSuCo 1 (ISC17)}, 2017.

\bibitem{Endo:1}
F.~Endo, D.~Courousse, and H.-P. Charles, ``Micro-architectural simulation of
  in-order and out-of-order arm microprocessors with gem5,'' in \emph{Embedded
  Computer Systems: Architectures, Modeling, and Simulation (SAMOS XIV), 2014
  International Conference on}, 2014, pp. 266--273.

\bibitem{Endo:2}
F.~A. Endo, D.~Courouss{\'e}, and H.-P. Charles, ``Micro-architectural
  simulation of embedded core heterogeneity with gem5 and {McPAT},'' in
  \emph{Proceedings of the Workshop on Rapid Simulation and Performance
  Evaluation: Methods and Tools}, 2015.

\bibitem{Asanovic06thelandscape}
K.~Asanovic, R.~Bodik, B.~C. Catanzaro, J.~J. Gebis, P.~Husbands, K.~Keutzer,
  D.~A. Patterson, W.~L. Plishker, J.~Shalf, S.~W. Williams \emph{et~al.},
  ``The landscape of parallel computing research: A view from {Berkeley},''
  Technical Report UCB/EECS-2006-183, EECS Department, University of
  California, Berkeley, Tech. Rep., 2006.

\bibitem{Hardkernel}
{ODROID-XU3}, http://www.hardkernel.com, [Accessed: May-2018].

\bibitem{Scalasca}
{Scalasca}, http://www.scalasca.org/, [Accessed: May-2018].

\bibitem{Vampir}
{Vampir - Performance Optimization}, https://www.vampir.eu/, [Accessed:
  May-2018].

\bibitem{scalrep}
F.~Desprez, G.~Markomanolis, and F.~Suter, ``Evaluation of profiling tools for
  the acquisition of time independent traces,'' INRIA, Tech. Rep., 2013.

\bibitem{6531079}
H.~Yun, G.~Yao, R.~Pellizzoni, M.~Caccamo, and L.~Sha, ``Memguard: Memory
  bandwidth reservation system for efficient performance isolation in
  multi-core platforms,'' in \emph{2013 IEEE 19th Real-Time and Embedded
  Technology and Applications Symposium (RTAS)}, April 2013, pp. 55--64.

\end{thebibliography}
}
%

\begin{IEEEbiography}
[{\includegraphics[width=1in,height=1in,clip,keepaspectratio]{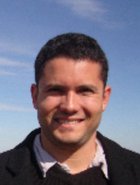}}]
{Dr. Florent Bruguier} received the M.S. and Ph.D. degrees in microelectronics from the University of Montpellier, France, in 2009 and 2012, respectively. From 2012 to 2015, he was a Scientific Assistant with the Montpellier Laboratory of Informatics, Robotics, and Microelectronics, University of Montpellier. Since 2015, he is a Permanent Associate Professor. He has co-authored over 30 publications. His research interests are focused on self-adaptive and secure approaches for embedded and high-performance systems.
\end{IEEEbiography}

\begin{IEEEbiography}[{\includegraphics[width=1in,height=1in,clip,keepaspectratio]{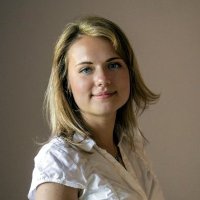}}]{Dr. Anastasiia Butko}
is a Postdoctoral Fellow in the Computational Research Division at Lawrence Berkeley National Laboratory (LBNL), CA. Her research interests lie in the general area of computer architecture, with particular emphasis on high-performance computing, emerging and heterogeneous technologies, associated parallel programming and architectural simulation techniques. 
Dr. Butko received her MSc. Degree in Microelectronics from UM2, France and MSc and BSc Degrees in Digital Electronics from NTUU "KPI", Ukraine. 
\end{IEEEbiography}

\begin{IEEEbiography}
[{\includegraphics[width=1in,height=1in,clip,keepaspectratio]{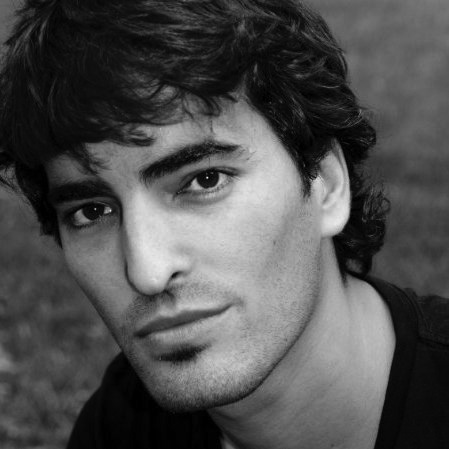}}]
{Dr. David Novo} received the M.S. degree from the Universitat Autonoma de Barcelona (UAB), Spain, in 2005, and the Ph.D. in Engineering from the KU Leuven, Belgium, in 2010. From 2010 to 2016, he was a postdoctoral researcher in the Processor Architecture Laboratory at EPFL (5 years), Switzerland, and in the Adaptive Computing group at LIRMM (1 year), France. Since January 2017, he is a tenured full-time CNRS research scientist at LIRMM. His research interests include hardware and software techniques for increasing computational efficiency in next-generation digital computers.
\end{IEEEbiography}

\begin{IEEEbiography}
[{\includegraphics[width=1in,height=1in,clip,keepaspectratio]{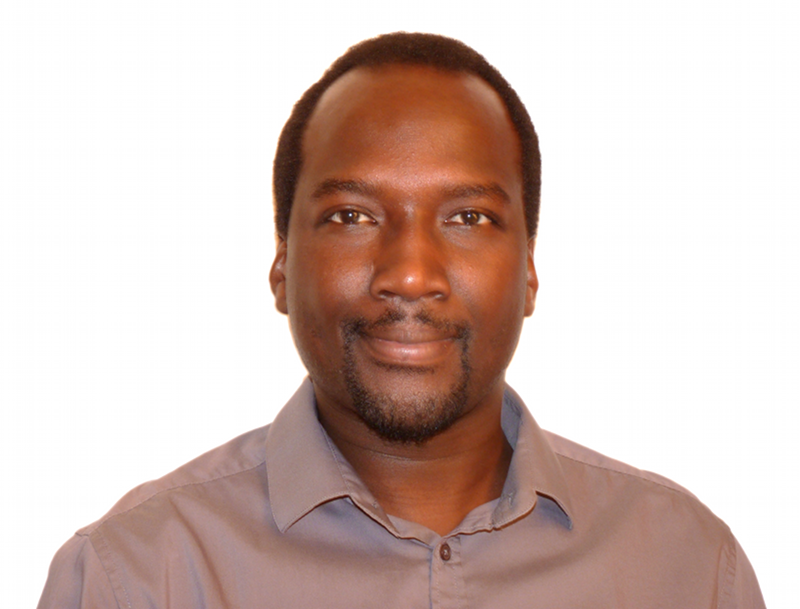}}]
{Dr. Abdoulaye Gamati\'{e}} is currently a CNRS Senior Researcher in the Microelectronics department of the LIRMM laboratory (Montpellier - France). His research activity focuses on the design of energy-efficient multicore/multiprocessor architectures for embedded and high-performance computing. He has also a long experience in the formal design of safety-critical embedded systems. He has been involved in several collaborative international projects with both academic and industrial partners. He co-authored more than 50 articles in peer-reviewed journals and international conferences. 
\end{IEEEbiography}

\begin{IEEEbiography}
[{\includegraphics[width=1in,height=1in,clip,keepaspectratio]{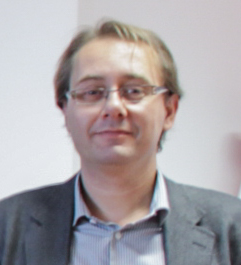}}]
{Dr. Gilles Sassatelli} is a CNRS senior scientist at LIRMM, a CNRS-University of Montpellier academic research unit with a staff of over 400. He is vice-head of the microelectronics department and leads a group of 20 researchers working in the area of smart embedded digital systems. He has authored over 200 peer-reviewed papers and has occupied key roles in a number of international conferences. Most of his research is conducted in the frame of international EU-funded projects such as the DreamCloud and Mont-Blanc projects.
\end{IEEEbiography}

%
%




\end{document}